\documentstyle[aps,12pt,epsf,manuscript]{revtex}

\newcommand{\bfsigma}{\mbox{\boldmath$\sigma$}}

\begin{document}
\tightenlines 
\draft
\preprint{KUNS-1581}
\title{	Nonequilibrium Weak Processes in Kaon Condensation I \\ 
--- Reaction rate for the thermal kaon process ---}

\author{Takumi Muto\thanks{p21mutot@cc.it-chiba.ac.jp}}	
\address{Department of Physics, Chiba Institute of Technology, 
2-1-1 Shibazono , Narashino, Chiba 275-0023, Japan}
\author{Toshitaka 
Tatsumi\thanks{tatsumi@ruby.scphys.kyoto-u.ac.jp}}   
\address{Department of Physics, Kyoto University,Kyoto 606-8502, 
Japan }
\author{Naoki Iwamoto\thanks{iwamoto@uoft02.utoledo.edu} }
\address{Department of Physics and Astronomy,
 The University of Toledo, Toledo, Ohio 43606-3390, U.S.A.}

\maketitle

\begin{abstract}
We investigate the thermal kaon process, 
in which kaons are thermally produced via nucleon-nucleon collisions. 
This process is relevant to nonequilibrium dynamics of kaon  condensation
inside neutron stars.  The reaction rates for these processes are calculated, 
and their temperature and 
density dependences are compared with those of other reaction rates. 
It is shown that the thermal kaon 
process is dominant over other relevant weak reactions 
throughout the nonequilibrium process, 
such as the kaon-induced Urca  
and the modified Urca reactions, and may control the entire 
evolution of the kaon condensate.  
The characteristic role of the soft and
hard kaons during the evolution is explained, and 
implications for astrophysical phenomena are briefly discussed. 
\end{abstract}

\newpage
\section{Introduction} 
\label{sec:intro}  
Modifications of properties of hadrons in a hot and/or dense 
nuclear medium have been an important subject in nuclear physics. 
 Specifically, kaon condensation has attracted much interest 
as a possible new hadronic phase in the context of 
astrophysics and heavy-ion physics\cite{kn86,t95,l96,blr98}. 
This condensate represents a Bose-Einstein condensation (BEC) 
of the negatively charged  kaons ($K^-$), driving
force for which comes mainly from  the $s$ wave kaon-nucleon ($KN$) 
attractions, i.e., the scalar interaction (the $KN$  sigma term)  
and the vector interaction (the Tomozawa-Weinberg term)\cite{t95,l96}. 
Kaon condensation may also be characterized 
as a macroscopic appearance of strangeness,  
which may have a close connection to strange matter. 
Based on experiments, the nature of attractive in-medium
antikaon-nucleon  interaction has recently been studied extensively in
antikaon-nucleon  scattering\cite{ww97,l98}, 
in kaonic atoms\cite{bfg97}, and in kaon-antikaon production via 
relativistic nucleus-nucleus collisions\cite{b97,l99,slk99}. 
In particular, the experimental  data 
for kaon-antikaon subthreshold production 
indicate a substantially reduced in-medium antikaon effective mass,  
which agrees with theoretical predictions\cite{l99}.  Thus, the basic
theoretical idea of kaon condensation seems to be  gaining experimental
support. 

Most of the discussions about kaon condensation so far 
have been focused on 
equilibrium properties of dense matter at low temperature:  
The appearance of a kaon-condensed phase in neutron stars 
would lead to softening of the  equation of state (EOS)\cite{fmmt96}. 
Its presence also brings about rapid cooling  
of neutron stars\cite{t88,fmtt94}. 
Recently, studies have been made of the possibility 
of a delayed collapse of protoneutron stars 
to low-mass black holes\cite{bb94,pbpelk97,bst96,kj95,p98}, 
possibly caused by the softening of EOS
by hadronic phase transitions during the evolution of protoneutron
stars.  In this case kaon condensation
should appear  in the high temperature and/or neutrino trapping 
situations. The equation of state for isothermal and isentropic cases 
and the properties of protoneutron stars with kaon condensation have been 
studied on the basis of chiral symmetry\cite{ty98,ty99}. 
However, the formation of a kaon condensate from    
noncondensed matter and its evolution involve  
{\it weak reaction processes} that change strangeness, 
so that it takes some time for chemical equilibrium to be reached: 
the typical time scale of the weak reactions is much longer       
than those for the strong and electromagnetic 
interactions, and is also different from 
the typical time scale for gravitational collapse 
(of order of a millisecond)  and that 
for deleptonization and initial cooling (of order of a second).  
Therefore, 
the nonequilibrium weak processes may have an important effect 
on the dynamical evolution of kaon-condensed stars 
just after supernova explosions. 

 Basic concept of nucleation in BEC 
is applied to a wide range of systems 
besides kaon condensation\cite{gss95}.   Concerning the nonequilibrium
dynamics and the formation time of  BEC, theoretical works have been
carried out   for a weakly interacting Bose gas\cite{st95},  prior to the first
experimental achievements of BEC in alkali atoms\cite{a95}.
In atomic gases, the energy of the atoms is little affected 
by the weakly repulsive interaction between the atoms, 
and the same interaction controls the onset and the build-up time 
scales of the condensate  with the number of atoms fixed.
On the other hand, the system of kaons and nucleons (which we call
the $KN$ system) is unique  in the following respects:
The microscopic quantities in the $KN$ system, 
such as the energies of the particles, 
are determined by the strong and
electromagnetic  interactions between the kaons, baryons and leptons. In
particular,  the kaons couple to the nuclear medium via strong interactions, 
and the kaon excitation energy is reduced primarily 
by $s$-wave $KN$ attractive interactions. 
The time scales for these interactions are much shorter 
than those of weak interactions, so that 
the $KN$ system is likely to be in thermal equilibrium 
during the chemical-nonequilibrium processes brought about by the
weak  reactions.   
As a result, the chemical-nonequilibrium state evolves adiabatically, 
adjusting to the gradual change of the chemical composition,   
which is determined by the kinetic equations induced 
by the weak reactions. 
The relevant such reactions are given by  
\begin{mathletters} \label{kt}
\begin{eqnarray}
& & n+n \rightarrow  n+p+K^- , \label{fkt} \\
 & &n+p+K^- \rightarrow n+n  \ ,    \label{bkt} 
\end{eqnarray}
\end{mathletters} 
where the presence of a spectator neutron is necessary in order to 
satisfy the kinematical conditions in a degenerate Fermi system. 
(We call these reactions thermal kaon 
process, and denote it as KT. )  
Other weak reactions  are given by the kaon-induced Urca process
(KU)\cite{t88},
\begin{mathletters}\label{ku}
\begin{eqnarray}
& & n(p)\rightarrow n(p) +e^- +\bar\nu_e, \label{fku} \\
& & n(p)+e^- \rightarrow n(p)+\nu_e \ , \label{bku}
\end{eqnarray}
\end{mathletters}
and the modified Urca process (MU)\cite{fm79,m87}, 
\begin{mathletters}\label{mu}
\begin{eqnarray}
& & n+n\rightarrow n+p +e^- +\bar\nu_e,\label{fmu} \\
& & n+p+e^- \rightarrow n+n+\nu_e \ . \label{bmu}
\end{eqnarray}
\end{mathletters}
Alternatively, the KU process (\ref{ku}) may be written 
symbolically as 
$n(p)+\langle K^-\rangle\rightarrow n(p)+e^- +\bar\nu_e$, 
$ n(p)+e^- \rightarrow n(p)+\langle K^-\rangle+\nu_e $, 
in terms of the condensate $\langle K^-\rangle$. It is to be noted that 
the KU process is  
operative only in the presence of a condensate, so that, 
in the noncondensed (normal) phase, only the KT and the MU processes 
proceed. 
Thus kaons, as quasiparticles whose properties are modified by 
 the medium effects, are emitted and absorbed through the weak reactions
(\ref{kt})  and (\ref{ku}), and the total number of kaons 
(and the corresponding strangeness in the medium)
changes  during the nonequilibrium weak processes. It is these weak
reactions  that determine the characteristic time scales of the system. 
\footnote{Concerning strangeness production, 
 some authors considered strange quark matter, 
where the nonleptonic weak reactions,
$u+d\rightleftharpoons s+u$, together with  the semileptonic ones, $d
(s)\rightarrow u+e^-+\bar\nu_e$, 
$u+e^- \rightarrow d (s)+\nu_e$, have been taken into 
account\cite{gps96}. 
One can see a similarity in these weak reactions between 
strange matter and  kaon condensation. }

In \cite{mti97}, we have discussed the nonequilibrium processes  
in a kaon condensate by taking into account only 
the KU and the MU reactions, and assuming the presence of 
a small seed of condensate. 
However, in thermal equilibrium, the onset of 
condensation can occur naturally without a seed when  
 the KT reactions are included: specifically, 
once the thermal kaon density is saturated, kaons are created 
thermally and then converted to a condensate. 
In a series of our works, 
we take into account the thermal kaon process KT as well as 
 KU and MU, and consider the effects of thermal kaons 
on the kinetics of kaon condensation 
during the whole nonequilibrium process.  
In this paper (paper I), we mainly consider the reaction rates for 
the thermal kaon process (\ref{kt}). We present a unified 
description of the excitation energy of kaons and the weak 
interactions based on chiral symmetry.  
In a subsequent paper (paper II)\cite{mti98-2}, 
the time dependent behavior of the physical quantities, such as 
the amplitude of the condensate, number densities of the chemical 
species, will be discussed, based on the results of this paper. 

This paper is organized as follows: 
The expressions for the reaction rates 
are obtained in Sec.\ref{sec:react}. 
Sec.\ref{sec:eos} outlines the physical properties of 
the nonequilibrium process.  
In Sec.\ref{sec:result}, the numerical 
results are presented, and temperature and the density 
dependences of these reaction rates are compared 
with other reaction rates. 
Summary and concluding remarks are given in Sec.\ref{sec:summary}. 
The expressions for other weak reaction rates for  KU 
and MU are outlined in Appendix A. 

\section{Formulation}
\label{sec:react}

\subsection{Brief survey of the chiral symmetry approach}
\label{ssec:chiral}

We first give an outline of the kaon-condensed state on the basis of 
chiral symmetry. In the framework of the $SU_L(3)\times SU_R(3)$ current 
algebra and PCAC, the $s$-wave $K^-$ condensed state,  
$|K^-\rangle$, is generated by a chiral rotation of 
the normal state $|0\rangle$ 
as $|K^-\rangle=\hat U_K |0\rangle$,  
with the unitary operator $\hat U_K$ given by 
\begin{equation} 
\hat U_K\equiv\exp(i\mu_K t Q) \exp(i\theta Q_4^5) \ , 
\label{uk}
\end{equation} 
where $\mu_K$ is the kaon chemical potential, $\theta$ the chiral 
angle which represents the order parameter of the system, and $Q$
($Q_4^5$) the electromagnetic charge operator (the axial charge
operator).  The classical
$K^-$ field is  then 
given as 
\begin{equation} 
\langle K^- \rangle\equiv \langle K^- |\hat K^-|K^-\rangle
=\langle 0|\hat U_K^{-1}\hat K^- \hat U_K|0\rangle
=\frac{f}{\sqrt 2}\sin\theta \exp(-i\mu_K t) \ , 
\label{kfield} 
\end{equation} 
where $f$(=93 MeV) is the meson decay constant.

\subsection{Reaction rate for thermal kaon process}
\label{ssec:rrate}

\begin{figure}[t]
\centerline{
\epsfxsize=0.6\textwidth\epsffile{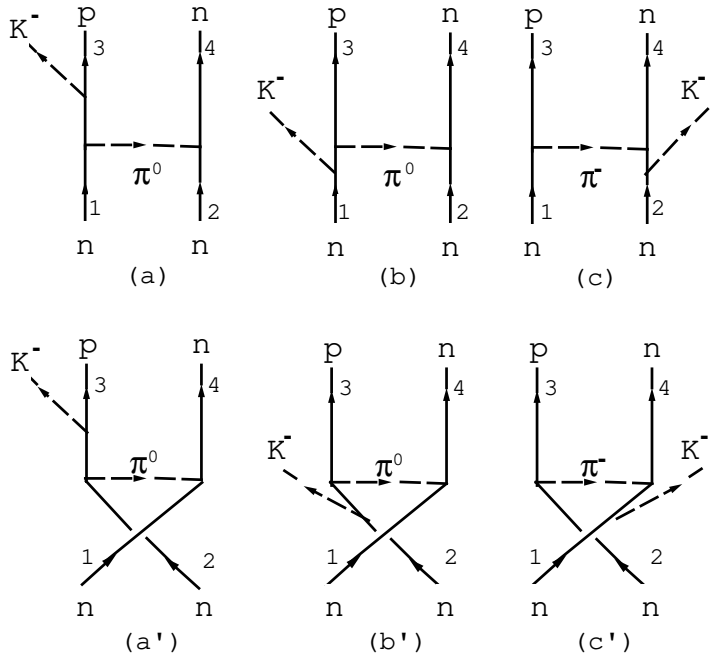}}
\caption{\footnotesize The lowest-order diagrams 
for the reactions (\ref{fkt}) that produce the thermal kaons. }
\label{fig1}
\end{figure}

Here we consider the kaon-producing (forward) reactions (\ref{fkt}).
\footnote{
Throughout this paper, we use the units in which 
$\hbar=k_{\rm B}=1$.} 
The lowest-order diagrams are 
given in Fig.\ref{fig1}, where $a-c$ represent the  
direct diagrams and $a' -c'$ the  exchange diagrams. 
In Fig.\ref{fig1}, lines 1 and 2 denote the neutrons in the initial 
state, and lines 3 and 4, the proton and the neutron in the final 
state, respectively. The scattering of two nucleons is 
treated in terms of the one-pion exchange potential (OPEP) 
which represents the long-range nuclear interaction.  
The medium and short-range interactions, 
which are mediated by heavier mesons, 
are  neglected here.\footnote{
We simply follow the treatment for the nuclear interaction adopted in
ref.\cite{m87}.}  

 The reaction rate per unit volume is given in terms of 
the transition rate $W_{\rm fi}$ by 
\begin{equation} 
 \Gamma^{(\rm KT-F)}=\frac{2\pi}{(2\pi)^{15}}
 \int d^3p_1 \int d^3p_2 \int d^3p_3 \int d^3p_4 
 \int d^3p_K \delta(E_{\rm f}-E_{\rm i})W_{\rm f i}S \ ,
 \label{rrfkt}
\end{equation}
where $S$ is the statistical factor,  
 $\displaystyle S=\frac{1}{s}n({\bf p_1})n({\bf p_2})[1-n({\bf p_3})]
 [1-n({\bf p_4})][1+f_{K^-}({\bf p}_K)]$, with the symmetry 
 factor $s$(=2), $n({\bf p_i})$ ($i=1-4$) the Fermi-Dirac distribution 
 function for the nucleon , and $f_{K^-}({\bf p}_K)$ 
 the Bose-Einstein distribution function for the $K^-$.   
 The nucleons and the kaons are 
 assumed to be in thermal equilibrium.  
  
\subsection{Matrix elements}
\label{subsec:matrix}

The transition rate $W_{\rm f i}$ is written as 
\begin{equation}
W_{\rm f i}=(2\pi)^3\delta^{(3)}({\bf p_3}+{\bf p_4}+{\bf p}_K
-{\bf p_1}-{\bf p_2})\sum_{\text{spins}}|M|^2 \ , 
\label{tm}
\end{equation}
where $|M|^2$ is the squared matrix element, and the 
summation is over the initial and final nucleon spins. 

 The effective weak Hamiltonian is of the current$-$ 
current interaction type,  \\ 
$\displaystyle H_{\rm W}={G_F\over \sqrt 2}
 J_{\rm h}^\mu\cdot J^\dagger_{{\rm h}\mu } +\text{h.c.}$ with 
the charged hadronic current,  
$J_{\rm h}^\mu=\cos\theta_C(V_{1+i2}^\mu -A_{1+i2}^\mu)
+\sin\theta_C(V_{4+i5}^\mu -A_{4+i5}^\mu) $, 
where $\theta_c(\simeq 0.24)$ is the Cabibbo angle,  
and $V_a^\mu$ and $A_a^\mu$ are the vector and axial vector currents, 
respectively. 

 In the kaon-condensed state $|K^-\rangle$, 
the matrix elements must be evaluated 
 from the transformed Hamiltonian\cite{t88,fmtt94}  
\begin{equation}
 {\widetilde H}_{\rm W}=\hat U_K^{-1}H_{\rm W}\hat U_K 
={G_{F}\over \sqrt 2}
{\widetilde J}_{\rm h}^\mu{\widetilde J}_{{\rm h}\mu}^\dagger 
+\text{ h.c.} \ ,
\label{th}
\end{equation}
where the effective hadronic current ${\widetilde J}_{\rm h}^\mu$ 
is given in a model-independent form by way of current algebra: 
\begin{eqnarray}
 {\widetilde J}^\mu_{\rm h}
&=&\hat U_K^{-1}J^\mu_{\rm h}\hat U_K \cr
&=&e^{-i\mu_K t}\Big\lbrack\cos\theta_c\bigg\{
(V_{1+i2}^\mu-A_{1+i2}^\mu)\cos(\theta/2)
+i(V_{6-i7}^\mu-A_{6-i7}^\mu)\sin(\theta/2)\bigg\} \cr
&+&\sin\theta_c\bigg\{(V_4^\mu-A_4^\mu)+i\cos\theta(V_5^\mu-A_5^\mu)
-{i\over 2}\sin\theta\Big(V_3^\mu-A_3^\mu+
\sqrt 3(V_8^\mu-A_8^\mu)\Big)\bigg\}\Big\rbrack \ .  
\label{tjh}
\end{eqnarray} 
Then the relevant isospin-changing strangeness-conserving 
 nucleon current can be read off from (\ref{tjh}) 
in the nonrelativistic form as 
 \begin{equation}
\widetilde{J_N}^\mu=e^{-i\mu_K t}\cos\theta_C\cos{\theta\over 2}
\chi^\dagger (p)(\delta^\mu_0-g_A\delta^\mu_i \sigma^i)\chi (n) 
\label{tjn}
\end{equation} 
with the two-component Pauli spinor 
$\chi(N)$ for the nucleon and $g_A$(=1.25) the axial-vector coupling
strength. The kaon sector in the hadronic current is implemented in 
the isospin and strangeness-changing axial vector 
terms in (\ref{tjh}), and may be written as
\begin{equation}
\widetilde{J_K}^\mu=-e^{-i\mu_K t}\sin\theta_C
\Big(\cos^2{\theta\over 2}j_{K^-}^\mu+
\sin^2{\theta\over 2}j_{K^+}^\mu\Big) 
\label{tjk}
\end{equation}
with $j_{K\pm}^\mu\equiv A_4^\mu\mp iA_5^\mu $.
In (\ref{tjk}), the first term in the bracket contributes 
to the thermal $K^-$ process. 
The kaon current $j_{K^\pm}^\mu$ may be determined  from 
the PCAC ansatz, $\partial_\mu j_{K^\pm}^\mu=\sqrt{2}f m_K^2K^\pm$, 
\footnote{
Generally speaking,  the kaon current consists of a free kaon part and  
a kaon-baryon interaction part.}
so that we take    
\begin{equation}
j_{K^\pm}^\mu=\sqrt{2}f\partial^\mu K^\pm
=\pm i\sqrt{2}f\frac{(p_{K^\pm})^\mu}{\sqrt{2\omega_\pm}}
\exp(\pm ip_{K^\pm} x) \ ,
\label{kc}
\end{equation}
where $(p_{K^\pm})^\mu=(\omega_\pm,{\bf p}_{K^\pm})$ is 
the kaon four-momentum with energy $\omega_\pm$. 
In a nuclear medium, the dispersion
relations for the kaons, 
$\omega_\pm({\bf p}_{K^\pm})$, differ from those in vacuum. 
See Sec.\ref{sec:eos} for the explicit expression for $\omega_\pm$.

For the pion-nucleon interaction, we take the pseudovector -coupling  
Lagrangian:
\begin{equation}
{\cal L}_{\pi NN}=\widetilde f\bar\Psi\gamma^\mu\gamma_5{\bf 
\tau}_a\Psi\partial_\mu\varphi_a + \text{h.c.}\ ,
\label{pnn}
\end{equation}
where $\widetilde f\equiv f_{\pi NN}/m_\pi$, with $f_{\pi NN}$ 
the $\pi NN$ coupling constant, $m_\pi$(=140 MeV) the pion mass,  
$\Psi$ the nucleon isodoublet, and $\varphi_a$ the pion isotriplet. 

The diagrams in Fig.1 are similar to those 
in the modified Urca (MU) process (\ref{mu}). 
The only difference between these processes 
is that a lepton pair $e^-$$\bar\nu_e$ in MU is replaced 
by a kaon in KT. Thus, the matrix element for KT 
can be obtained by a replacement of the $npe^-\bar\nu_e$ 
vertex factor in MU by a $npK^-$ vertex factor.
In addition, the KT and MU reactions differ 
in the statistical factor $S$ [ see Eq.(\ref{rrfkt}) for KT ], 
as a result of the different statistics of the participating particles. 
 From (\ref{th})$-$(\ref{kc}), the vertex factor can be 
read as $\displaystyle\frac{G_F}{\sqrt 2}\cos\theta_C K_\mu 
(\delta^\mu_0-g_A\delta^\mu_i \sigma^i) $,  
where  
\begin{equation}
 K_\mu\equiv-if\sin\theta_C
\frac{(p_K)_\mu}{\omega_-^{1/2}}\cos^3{\theta\over 2}\ . 
\label{kmu}
\end{equation}

The remaining assignments in the matrix elements of Fig.1 are made  
with reference to Friman and Maxwell's result 
for the  MU process\cite{fm79,m87}:

\noindent (i) A factor $-\widetilde f
\bfsigma\cdot{\bf k}$ for
the $nn\pi^0$ vertex,  a factor $\widetilde f\bfsigma
\cdot{\bf k}$ for the $pp\pi^0$ vertex, and a factor 
$\sqrt{2}\widetilde f\bfsigma\cdot{\bf k}$ for the
$np\pi^-$ vertex,  where ${\bf k}$ is the pion momentum. 

\noindent (ii) A factor $-1/({\bf k}^2+m_\pi^2)$ 
for the pion propagator . 

\noindent (iii) A nucleon propagator $iG$ 
at the nucleon internal line, which,  
when the kaon couples to the outgoing (incoming) nucleon, 
is given by 
$\displaystyle i/(E_p+\omega_- -E_n)$
( $\displaystyle i/(E_n-E_p-\omega_-)$ ), 
where $E_N$ is the single particle energy of the nucleon. 
The nucleon propagator is further expanded 
in powers of the inverse nucleon mass, and only the lowest order 
term is retained, 
so that $iG\sim i/\omega_-$ ($ iG\sim -i/\omega_- $ ). 

Here we summarize the matrix element for each diagram for the 
kaon-producing KT process (\ref{fkt}).
For the direct diagrams (a)$-$(c), one obtains
\begin{mathletters}\label{dm}
\begin{eqnarray}
 M^{(\rm a)}(1,2,3,4;l) &=&
 -i\frac{G_F}{\sqrt 2}\cos\theta_C 
 K_\mu\widetilde f^2\frac{1}{\omega_-}\frac{1}{{\bf 
 l}^2+m_\pi^2}
 \Big\{\chi_4^\dagger(\bfsigma\cdot{\bf l})\chi_2\cdot 
 \chi_3^\dagger(\delta^\mu_0-g_A\delta_i^\mu\sigma^i)
 (\bfsigma\cdot{\bf l})\chi_1 \Big\} \ ,\cr
 & &\hspace{10.0cm} \label{ma} \\
  M^{(\rm b)}(1,2,3,4;l) 
 & =&-i\frac{G_F}{\sqrt 2}\cos\theta_C 
 K_\mu\widetilde f^2\frac{1}{\omega_-}\frac{1}{{\bf 
 l}^2+m_\pi^2}
 \Big\{\chi_4^\dagger(\bfsigma\cdot{\bf l})\chi_2\cdot 
 \chi_3^\dagger(\bfsigma\cdot{\bf l})
 (\delta^\mu_0-g_A\delta_i^\mu\sigma^i)
 \chi_1 \Big\} \ , \cr
 & &\hspace{10.0cm}\label{mb} \\
  M^{(\rm c)}(1,2,3,4;k) 
 & =&i\frac{G_F}{\sqrt 2}\cos\theta_C 
 K_\mu \cdot 2\widetilde f^2\frac{1}{\omega_-}\frac{1}{{\bf 
 k}^2+m_\pi^2}
 \Big\{\chi_4^\dagger(\bfsigma\cdot{\bf k})
 (\delta^\mu_0-g_A\delta_i^\mu\sigma^i)\chi_2\cdot 
 \chi_3^\dagger(\bfsigma\cdot{\bf k})\chi_1 \Big\} \ , \cr
 & &\hspace{10.0cm}\label{mc}
\end{eqnarray}
\end{mathletters}
where $k= p_1-p_3$, $l= p_4-p_2$.

For the exchange diagrams (a')$-$(c'), one obtains
\begin{mathletters}\label{em}
\begin{eqnarray}
 M^{(\rm a')}&=& -M^{(\rm a)}(2,1,3,4;-k')  \cr
 &=& i\frac{G_F}{\sqrt 2}\cos\theta_C 
 K_\mu\widetilde f^2\frac{1}{\omega_-}\frac{1}{{\bf 
 k'}^2+m_\pi^2}
 \Big\{\chi_4^\dagger(\bfsigma\cdot{\bf k'})\chi_1\cdot 
 \chi_3^\dagger(\delta^\mu_0-g_A\delta_i^\mu\sigma^i)
 (\bfsigma\cdot{\bf k'})\chi_2 \Big\} \ ,\cr
 & &\hspace{10.0cm}\label{map} \\
 M^{(\rm b')}&=&-M^{(\rm b )}(2,1,3,4;-k') \cr
 & =& i\frac{G_F}{\sqrt 2}\cos\theta_C 
 K_\mu\widetilde f^2\frac{1}{\omega_-}\frac{1}
 {{\bf k'}^2+m_\pi^2}
 \Big\{\chi_4^\dagger(\bfsigma\cdot{\bf k'})\chi_1\cdot 
 \chi_3^\dagger(\bfsigma\cdot{\bf k'})
 (\delta^\mu_0-g_A\delta_i^\mu\sigma^i)
 \chi_2 \Big\} \ , \cr
 & & \hspace{10.0cm}\label{mbp} \\
M^{(\rm c')}&=&- M^{(\rm c)}(2,1,3,4;-l') \cr
 & =&-i\frac{G_F}{\sqrt 2}\cos\theta_C 
 K_\mu \cdot 2\widetilde f^2\frac{1}{\omega_-}\frac{1}{{\bf 
 l'}^2+m_\pi^2}
 \Big\{\chi_4^\dagger(\bfsigma\cdot{\bf l'})
 (\delta^\mu_0-g_A\delta_i^\mu\sigma^i)\chi_1\cdot 
 \chi_3^\dagger(\bfsigma\cdot{\bf l'})\chi_2 \Big\} \ , \cr
 & & \hspace{10.0cm}\label{mcp}
\end{eqnarray}
\end{mathletters}
where $k'=p_1-p_4$, $l'=p_3-p_2$.

From momentum conservation, ${\bf k}={\bf l}\pm {\bf p}_K$ and
${\bf k'}={\bf l'}\pm {\bf p}_K$. Since the nucleons are degenerate,
only those near the Fermi surfaces contribute to the
relevant phase space  of the reactions. 
Thus, the momenta for the nucleons
in these  relations may be replaced by the Fermi momenta, i.e., 
${\bf p}_F (n)$ for ${\bf p}_1$, ${\bf p}_2$, and ${\bf p}_4$, 
 and ${\bf p}_F (p)$  for ${\bf p}_3$. 
 As we will see in
Sec.\ref{sec:result}, the proton Fermi momentum 
in the noncondensed state is estimated as 
$|{\bf p}_F(p)|=$ 260$-$290 MeV for the typical baryon number densities 
$n_B$=0.55$-$ 0.70 fm$^{-3}$, whereas the corresponding neutron Fermi
momentum  is $|{\bf p}_F(n)|$=470$-$510 MeV. In addition, 
the typical value of the momentum of the thermal kaons which mainly
contributes to the reactions is 120$-$ 180 MeV. Thus, both
$|{\bf p}_F(p)|$ and $|{\bf p}_K|$ are small compared  
with $|{\bf p}_F(n)|$ at least in the noncondensed stage.  Therefore we, 
hereafter, make an approximation by
omitting the proton and the kaon  momenta 
in the momentum conservation, 
and we put ${\bf k}= {\bf l}\simeq {\bf p}_F(n)$ and
${\bf k'}={\bf l'}\simeq -{\bf p}_F(n)$.
Within this approximation, one obtains 
\begin{eqnarray}
 M^{(\rm a)}+M^{(\rm b)}+M^{(\rm c)}
&=&2g_A\frac{G_F}{\sqrt{2}}\cos\theta_C\widetilde 
f^2\frac{1}{\omega_-}
\frac{1}{|{\bf p}_F(n)|^2+m_\pi^2}K_\mu\delta^\mu_i \cr
&\cdot&\Big\lbrack \chi_3^\dagger\chi_1\cdot \chi_4^\dagger
(\bfsigma\cdot {\bf k})\chi_2(ik^i)
-\chi_3^\dagger(\bfsigma\cdot {\bf k})\chi_1
\Big\{\chi_4^\dagger\chi_2(ik^i)
-\chi_4^\dagger(\bfsigma\times{\bf k})^i\chi_2\Big\}\Big\rbrack 
\label{mabc}
\end{eqnarray}
and 
\begin{eqnarray}
 M^{(\rm a')}+M^{(\rm b')}+M^{(\rm c')}
&=&2g_A\frac{G_F}{\sqrt{2}}\cos\theta_C\widetilde 
f^2\frac{1}{\omega_-}
\frac{1}{|{\bf p}_F(n)|^2+m_\pi^2}K_\mu\delta^\mu_i \cr
&\cdot& \Big\lbrack \Big\{\chi_4^\dagger\chi_1(ik'^i)
-\chi_4^\dagger(\bfsigma\times{\bf k'})^i\chi_1\Big\}
\chi_3^\dagger(\bfsigma\cdot {\bf k'})\chi_2
-\chi_4^\dagger(\bfsigma\cdot {\bf k'})\chi_1
\cdot\chi_3^\dagger\chi_2 (ik'^i)\Big\rbrack \ .
\label{mabcp}
\end{eqnarray}
It can be checked that the expressions for 
these summed matrix elements 
(\ref{mabc}) and (\ref{mabcp}) are consistent with 
those for the MU process obtained by Maxwell\cite{m87} 
by simple replacements of the weak lepton current 
$l_\mu$ and the energy $\omega$ of the lepton pairs 
in \cite{m87}, by $K_\mu$ and $\omega_-$, respectively.  

It is to be noted that the nucleon vector current part 
of the matrix element for the diagram (c) ( (c') ),where a charged pion
propagates  between the nucleons, is larger 
by a factor of two and is opposite in 
sign as compared with those for (a) and (b) [ (a') and (b') ], 
where a neutral pion propagates. [ cf. (\ref{dm}) and (\ref{em}).]
This is due to the differences in the isospin factors
 at the $\pi NN$ vertices and in the sign attached 
to the nucleon propagator (see (iii) ) between (c) and (a) or (b)
 [(c') and (a') or (b') ].
As a result, the contribution from the nucleon  
vector current vanishes after summing up  the matrix elements for the
three diagrams (a), (b) and (c)  [ (a'), (b') and (c') ],  
and only the axial-vector part, proportional to 
$g_A \delta^\mu_i$,  contributes, as seen in (\ref{mabc}) and (\ref{mabcp}). 
One finds that the $s$-wave condensed kaon cannot be 
directly emitted in place of the thermal kaon in the KT reaction. 
This is because the relevant kaon current has the form  
 $ j_{K^-}^\mu=\sqrt{2}f\partial^\mu\langle K^-\rangle
 =-i\mu_K f^2\sin\theta e^{-i\mu_K t}\cdot\delta^\mu_0 $, 
where the spatial component is zero, 
 so that the condensed-kaon current does not couple to the 
 nonrelativistic nucleon axial-vector  current 
[ cf. (\ref{mabc}) and (\ref{mabcp})]. 
 
By the use of (\ref{mabc}) and (\ref{mabcp}), 
the spin-summed squared matrix element 
$\displaystyle\sum_{\text{spins}}|M|^2$
, with $M=M^{(a)}+M^{(b)}+M^{(c)}+M^{(a')}+M^{(b')}+M^{(c')}$, is 
obtained as 
\begin{eqnarray}
\sum_{\text{spins}}|M|^2
&=&8\Big(g_A G_F{\widetilde f}^2f\sin\theta_C\cos\theta_C
\cos^3\frac{\theta}{2}\Big)^2\frac{1}{\omega_-^3}
\frac{1}{(|{\bf p}_F(n)|^2+m_\pi^2)^2}\cr
\cdot&\Bigg\lbrack&|{\bf p}_F(n)|^2\Big\{({\bf p}_K \cdot{\bf k})^2+
({\bf p}_K\cdot{\bf k'})^2+2|{\bf p}_K|^2 |{\bf p}_F(n)|^2\Big\} \cr
&+&
\Big\{5({\bf k}\cdot{\bf k'})({\bf p}_K\cdot{\bf k})
({\bf p}_K\cdot{\bf k'})-2|{\bf p}_F(n)|^2({\bf p}_K\cdot{\bf k'})^2
-2|{\bf p}_F(n)|^2({\bf p}_K\cdot{\bf k})^2 \cr
&-&|{\bf p}_K|^2 ({\bf k}\cdot{\bf k'})^2+\big\{{\bf p}_K
\cdot({\bf k}\times{\bf k'})\big\}^2\Big\}
\Bigg\rbrack \ .
\label{sssm}
\end{eqnarray}

\subsection{Phase space integrals}

 The phase space integrals appearing in (\ref{rrfkt})
for the reaction rate $\Gamma^{({\rm KT})}$ are separated into 
 radial and angular parts via the relation $\Gamma^{({\rm KT})}=AP$,
where
\begin{equation}
A\equiv\frac{1}{(2\pi)^{11}}\Bigg(\frac{8}{3}
g_AG_F{\widetilde f}^2f\sin\theta_C
\cos\theta_C\cos^3{\theta\over 2}
\frac{|{\bf p}_F(n)|^2}{|{\bf p}_F(n)|^2+m_\pi^2}\Bigg)^2 \ ,
\label{phasea}
\end{equation}
and 
\begin{equation}
P\equiv \int_0^\infty |{\bf p}_1|^2d|{\bf p}_1|\int_0^\infty 
|{\bf p}_2|^2d|{\bf p}_2|
\int_0^\infty |{\bf p}_3|^2d|{\bf p}_3|\int_0^\infty
|{\bf p}_4|^2d|{\bf p}_4|
\int_0^\infty |{\bf p}_K|^2d|{\bf p}_K|\delta(E_f-E_i)
\frac{|{\bf p}_K|^2}{\omega_-^3} S Q
\label{phasep}
\end{equation}
with
\begin{equation}
Q\equiv \int d\Omega_1\int d\Omega_2\int d\Omega_3
\int d\Omega_4\int d\Omega_K
\delta^{(3)}({\bf p}_3+{\bf p}_4+{\bf p}_K-{\bf p}_1-{\bf p}_2) \ .
\label{phaseq}
\end{equation}
For those kaons which are the main contributors to the reaction rate, 
the momentum $|{\bf p}_K|$ is small enough to satisfy the
condition,  $|{\bf p}_K|+|{\bf p}_F(p)| <|{\bf  p}_F(n)|$. 
Under this condition, one  obtains
$Q\simeq 8(2\pi)^4|{\bf p}_F(p)|/(|{\bf p}_1||{\bf p}_2||{\bf p}_3||{\bf
p}_4|)$. 
 After substituting 
this result into (\ref{phasep}), we transform the variables for the 
nonrelativistic nucleons in $P$ by way of the relation
 $|{\bf p}_i|d|{\bf p}_i|=m_i^\ast  dE_i$ ($i=1- 4$) 
with $m_i^\ast$ the effective nucleon mass. 
By the further change of variables : $x_1=(E_1-\mu_n)/T$,
$x_2=(E_2-\mu_n)/T$, 
$x_3=-(E_3-\mu_p)/T$, $x_4=-(E_4-\mu_n)/T$, 
where  $\mu_i$ ($i=1-4$) is the chemical 
potential and $T$ the temperature, one obtains 
\begin{equation}
P=\frac{1}{6}(m_n^\ast)^3 m_p^\ast T^3\int_0^\infty 
|{\bf p}_K|^2 d|{\bf p}_K| \frac{[(\omega_- +\mu_p-\mu_n)/T]
\cdot\big\lbrack
(\omega_- +\mu_p-\mu_n)^2/T^2+4\pi^2\big\rbrack}
{e^{(\omega_- +\mu_p-\mu_n)/T}-1}
\frac{|{\bf p}_K|^2/\omega_-^3}{1-e^{-(\omega_- - \mu_K)/T}}  .
\label{pp}
\end{equation}
In obtaining (\ref{pp}), 
we have made the approximation 
(valid at low temperature), $-\mu_i/T\rightarrow-\infty$, 
for the lower limit of the four integrals with respect to $x_1 - x_4$,
and the following formula has been used \cite{bp91}:
\begin{eqnarray}
& &\int_{-\infty}^{\infty}dx_1\int_{-\infty}^{\infty}dx_2
\int_{-\infty}^{\infty}dx_3\int_{-\infty}^{\infty}dx_4
\delta(x_1+x_2+x_3+x_4-y)\frac{1}{1+e^{x_1}}\frac{1}{1+e^{x_2}}
\frac{1}{1+e^{x_3}}\frac{1}{1+e^{x_4}} \cr
&=&\frac{1}{6}\frac{y(y^2+4\pi^2)}{e^y-1} \ . 
\label{formula}
\end{eqnarray}
The value of $\mu_n$ ($\mu_p$) is taken to be 
200$-$300 MeV (100$-$200 MeV) at the relevant densities 
where the kaon condensation is realized. As a consequence of 
 the rapid convergence of the integrand in (\ref{formula}) 
for negative values of $x_i$ produced by 
the exponentials arising from the statistical factor 
$S$,  the low-temperature approximation should be valid for 
temperatures less than several tens of MeV. 

For the remaining integral in (\ref{pp}), we introduce a  
dimensionless variable $x\equiv |{\bf p}_K|/T$. 
Combining Eqs. (\ref{phasea}) and 
(\ref{pp}), and the definition of $ \Gamma^{\rm (KT-F)}$, 
one obtains the final expression  for $ \Gamma^{\rm (KT-F)}$ :
\begin{mathletters}\label{frrfkt}
\begin{eqnarray}
\Gamma^{\rm (KT-F)}(\xi^{\rm (KT)},T)&=&
\frac{512}{9(2\pi)^7}\Bigg(g_AG_F{\widetilde f}^2f\sin\theta_C
\cos\theta_C\cos^3{\theta\over 2}
\frac{|{\bf p}_F(n)|^2}{|{\bf p}_F(n)|^2+m_\pi^2}\Bigg)^2 |{\bf p}_F(p)|
\cr &\times&(m_n^\ast)^3m_p^\ast T^5 
I^{({\rm KT})}(\xi^{\rm (KT)},T) \label{frrfkta} \\
&=&(4.0\times 10^{30})\bigg({|{\bf p}_F(p)| \over m_\pi}\bigg)
\bigg({m_n^\ast\over m_N}\bigg)^3
\bigg({m_p^\ast\over m_N}\bigg) 
\cos^6{\theta\over 2}T_9^5 I^{\rm (KT)}(\xi^{\rm (KT)},T) \cr 
& &\hspace{8.0cm} ({\rm cm}^{-3}\cdot {\rm s}^{-1}) \ ,\label{frrfktb}
\end{eqnarray}
\end{mathletters}
where 
$T_9$ is the temperature in units of 10$^9$ K, 
$ \xi^{\rm (KT)} \equiv (\mu_K+\mu_p-\mu_n)/T$, 
and 
\begin{equation}
 I^{\rm (KT)}(u,T)=\int_0^\infty dx f(x;u,T) \ , 
\label{ikt}
\end{equation}
with 
\begin{equation}
f(x;u,T)
\equiv {1\over 6}\frac{x^4}{(\widetilde{\omega_-}(x)+\mu_K/T)^3}
\frac{(\widetilde{\omega_-}(x)+u)}{1-e^{-\widetilde{\omega_-}(x)}}
\frac{[(\widetilde{\omega_-}(x)+u)^2+4\pi^2]}{
e^{\widetilde{\omega_-}(x)+u}-1} 
\label{integrandf}
\end{equation}
where $\widetilde{\omega_-}(x)\equiv(\omega_- (x)-\mu_K)/T $. 

 The factor $T^5$ in (\ref{frrfkt}) comes from the energy integrals 
 for two incoming and two outgoing nucleons ($T^4$), together with  
the energy-conserving delta function ($T^{-1}$),     
the radial integral with respect to the kaon 
momentum ($T^3$), and the factor $|{\bf p}_K|^2/\omega_-^3$ in the
matrix  element ($T^{-1}$). 
The integral $I^{({\rm KT})}(\xi^{\rm (KT)},T)$
also depends on $T$, so that the reaction rate 
$\Gamma^{\rm (KT-F)}(\xi^{\rm (KT)},T)$ in general 
has a complicated temperature dependence. 
 
For the backward process (\ref{bkt}), the reaction rate 
$\Gamma^{\rm (KT-B)}(\xi^{\rm (KT)},T)$ can be given 
in a manner similar to the forward process. Noting that the statistical 
factor $S$ is replaced by  
$\displaystyle S'=\frac{1}{s}n({\bf p}_3)n({\bf p}_4)f_{K^-}({\bf p}_K)
[1-n({\bf p}_1)][1-n({\bf p}_2)]$, one obtains 
\begin{equation}
\Gamma^{\rm (KT-B)}(\xi^{\rm (KT)},T)
=e^{\xi^{\rm (KT)}}\Gamma^{\rm (KT-F)}(\xi^{\rm(KT)},T) 
\label{brrkt}
\end{equation}
in the low temperature approximation. 

The initial depletion of total strangeness is reflected in a negative 
value of $\xi^{\rm (KT)}$. Consequently, the 
 production rate of thermal kaons through the forward KT reaction 
is larger than the annihilation rate through the backward 
reaction, as seen from (\ref{brrkt}). Eventually, when the system 
reaches chemical equilibrium where $\xi^{\rm (KT)}$=0,  
both the forward and the backward reaction rates become equal. 

\section{EOS and Physical conditions}
\label{sec:eos}

In order to estimate the reaction rates, we must know 
for each chemical species the
values of  the physical quantities such as the chemical potential $\mu_i$ 
 and the number density $n_i$ ($i=n,p,e^-,K^-$), which depend on the
specific physical properties of the system.  
As the initial condition ($t$=0), 
 we take normal neutron-star matter
($\theta=0$) which is composed of nonrelativistic neutrons ($n$), 
protons ($p$), and ultrarelativistic  free electrons ($e^-$) with a baryon number density $n_B$. 
The value of $n_B$ is taken to be larger than the critical density 
for kaon condensation $n_B^C$.  The initial kaon chemical potential 
 $\mu_K^0$ has a large negative value 
[see the latter part of this section], 
while $\mu_e^0$ is positive. \footnote{The superscript `0' 
denotes the  initial value at $t=0$.}
Then the system is far from chemical equilibrium 
with $\mu_e^0\neq \mu_K^0$, and evolves into a chemically equilibrated
kaon-condensed phase.  The temporal evolution of the kaon condensation
proceeds  through a set of rate equations which are given by   
the nonequilibrium weak reactions\cite{mti98-2}. 

As for the  EOS, the thermodynamic potential $\Omega$ 
for the kaon-condensed phase is  constructed 
with the help of chiral symmetry.  
Here we basically follow  the result of \cite{ty98} for  $\Omega$. 
During the nonequilibrium processes, the time scales for the 
weak interactions are  much larger than those for the 
strong and the electromagnetic interactions 
which are responsible for thermal equilibration. 
Therefore, the physical quantities are determined adiabatically 
under the assumption that the system is in thermal equilibrium, 
adjusting to the gradual change of the chemical compositions 
given by the weak interactions.  Under this assumption, 
 the corresponding values of $\mu_i(t)$  and 
$n_i(t)$ are related to each other, $n_i(t)=-\partial
\Omega(t)/\partial \mu_i(t)$.  Here we should note that both the
zero-point and temperature fluctuations are simply neglected 
in the EOS except for the thermal contribution 
to the kaon number density, so that our expression
for the EOS  corresponds to the heavy-baryon limit in
ref.\cite{ty98} for the nucleons.  
On the other hand,  to make clear the role of  
thermal kaons on the kinetics of condensation, we explicitly 
take into account the number of thermal kaons. 
Thus the kaon number density
$n_K$ is written as a sum of  the condensed part 
$\zeta_K$ and the thermal part $n_K^T$, 
as $n_K(t)=\zeta_K(t)+n_K^T(t)$, where
\begin{equation}
\zeta_K(t)\equiv \langle K^-|{\hat S}|K^-\rangle
=\mu_K(t)f^2\sin^2\theta(t)
+\big(1-\cos\theta(t)\big)\bigg(n_p(t)
+{1\over 2}n_n(t)\bigg) 
\label{eqstrangeness}
\end{equation}
with the strangeness operator, 
${\hat S}\equiv 2({\hat Q}-{\hat I}_3-{\hat B})$,  
\begin{equation}
 n_K^T(t)={1\over (2\pi)^3}\cos\theta\int d^3p_K f_{K}({\bf p}_K,t)
\label{nkt}
\end{equation}  
with 
\begin{equation}
f_{K}({\bf p}_K,t)=
{1\over e^{(\omega_-(t) -\mu_K(t))/T}-1}
-{1\over e^{(\omega_+(t) +\mu_K(t))/T}-1}\ ,
\label{eqdistribution}
\end{equation}
and the first and second terms in the RHS of (\ref{eqdistribution}) 
are the Bose-Einstein distribution functions of the $K^-$ 
and $K^+$ mesons, respectively. It is to be noted that there appears 
a  factor $\cos\theta$ in the number density $n_K^T(t)$, 
which reduces to the noncondensed form 
in the limit of $\theta\rightarrow 0$ ($\cos\theta\rightarrow
1$)\cite{ty98}.  

 In the condensed phase, 
 the expression for the $\omega_{\pm}({\bf p}_K)$ depends on how
 the fluctuations are incorporated\cite{ty98,te97}.  
 Here we adopt the result by Tatsumi and Yasuhira\cite{ty98}. 
The expression for $\omega_{\pm}({\bf p}_K)$ is then given by 
\begin{equation}
\omega_{\pm}({\bf p}_K)=\pm\Big\lbrace
b+\mu_K(\cos\theta-1)\Big\rbrace +\Big\lbrack {\bf
p}_K^2+(b^2+\widetilde{ m_K^{\ast 2}})
\Big\rbrack^{1/2} \ ,  
\label{cenergy}
\end{equation}
where $\displaystyle b\equiv \Big(n_p+\frac{1}{2}n_n\Big)/(2f^2)$ 
with $n_p$ ($n_n$) the proton (neutron) number density, 
and $\widetilde {m_K^{\ast 2}}\equiv m_K^{\ast 2}\cos\theta
=(m_K^2-n_B\Sigma_{\rm KN}/f^2)\cos\theta$, 
and $\Sigma_{\rm KN}$ is the $KN$ sigma term .  
The term $b$ stems from the $KN$ vector interaction 
(the Tomozawa-Weinberg term), and 
$m_K^{\ast 2}$ is the effective kaon mass squared which is reduced due to 
the $KN$ scalar interaction simulated by 
$\Sigma_{\rm KN}$\cite{fmmt96}. 

It may be seen from (\ref{cenergy}) that 
$\omega_-({\bf p}_K)\rightarrow \mu_K$ as ${\bf p}_K\rightarrow 0$,
with the  help of the classical field equation for
$\theta$,$\partial\Omega_K/\partial\theta=0$\cite{fmmt96,ty98}, 
\begin{equation}
\sin\theta\Big(\mu_K^2\cos\theta+2b\mu_K
-m_K^{\ast 2}\Big)=0 \ . 
\label{fieldeq}
\end{equation}
In particular, in the condensed phase ($\theta\neq 0$), the lowest
excitation energy of the thermal  kaons coincides 
with the kaon chemical potential.  This Goldstone-like nature 
originates from the spontaneous $V$-spin  symmetry breaking in the
condensed phase.  In the limit of $\theta=0$ in Eq.(\ref{cenergy}),  
one obtains an expression for the excitation energy of kaons 
in the normal phase. In this case, 
there is a gap between the lowest excitation energy 
$\omega_-({\bf p}_K=0)$ and $\mu_K$. 

For the EOS, the potential contribution 
of the symmetry energy,  
$V_{\rm sym}(n_B)$, is taken into account in addition to the 
nonrelativistic Fermi-gas energy for the nucleons.  
Following Prakash et al.\cite{pal88}, we take the density
dependence of $V_{\rm sym}(n_B)$ in the form     
\begin{equation}
V_{\rm sym}(n_B)=\Big\lbrack
S_0-(2^{2/3}-1)\frac{3}{5}\epsilon_{F,0}\Big\rbrack F(n_B) \ ,
\label{symmetry}
\end{equation}
where $S_0$(=30 MeV) is the empirical symmetry energy, 
$\epsilon_{F,0}$ is the Fermi energy in the symmetric nuclear matter 
at $n_0$(=0.16 fm$^{-3}$ ), the
standard nuclear matter density, and $F(n_B)$ is a function simulating 
the density dependence of $V_{\rm sym}(n_B)$. 
For simplicity, we take $F(n_B)=n_B/n_0$. 
Then the chemical potential difference between 
the proton and the neutron in the condensed state is given by 
\begin{eqnarray}
\mu_p(t)-\mu_n(t)&=&\frac{(3\pi^2 n_p(t))^{2/3}}{2m_N}-
\frac{(3\pi^2n_n(t))^{2/3}}{2m_N}+4V_{\rm sym}(n_B)\cdot
(n_p(t)-n_n(t))/n_B \cr
&-&\frac{1}{2}\mu_K(t)(1-\cos\theta(t))  \ ,
\label{ceppn}
\end{eqnarray}
where $n_p(t)$ ($n_n(t)$) is the proton (neutron) 
number density\cite{ty98}. 

With relevance to hot neutron stars at birth,  
we further assume that the
system is initially in $\beta$-equilibrium , i.e.,
$\mu_n^0=\mu_p^0+\mu_e^0$,  due to the rapid $\beta$-decay
reactions, $ n\rightarrow p+e^-+\bar 
\nu_e$, $ p+e^-\rightarrow n+\nu_e $, and that the initial 
total strangeness is almost zero  (with equal numbers 
of thermal $K^+$'s and $K^-$'s present).  The latter condition means 
$\mu_K^0=-b^0$ as a consequence of  Eqs.(\ref{nkt}), 
(\ref{eqdistribution}), and  (\ref{cenergy}) with $\theta=0$.  
With these initial conditions, one can obtain the number densities 
$n_i(t)$ from the rate equations, and then the chemical potentials
$\mu_i(t)$ and $\theta(t)$ from Eqs.(\ref{fieldeq}) and (\ref{ceppn})
together with the charge neutrality condition, 
\begin{equation}
n_p(t)=n_e(t)+n_K(t) \ . 
\label{neutral}
\end{equation}

 Hereafter, we take the values for the effective nucleon  mass
ratios which appear in the expressions for the reaction rates to be 
$m_p^\ast/m_N=m_n^\ast/m_N$=0.8 for simplicity. 
The value for the $KN$ sigma term is chosen to be
$\Sigma_{KN}$=300 MeV as  an example\cite{dn86}. 
The critical density $n_B^C$ is then estimated to be $n_B^C$=
0.485 fm$^{-3}$ ($\sim$3.0 $ n_0$). 

\section{Numerical results and discussion}
\label{sec:result}

In this section, we discuss the characteristic features of the 
forward KT reaction rate $\Gamma^{\rm (KT-F)}(\xi^{\rm (KT)},T)$. 
The reaction rate  
$\Gamma^{\rm (KT-F)}(\xi^{\rm (KT)},T)$ is evaluated 
in the following two typical stages in the nonequilibrium process:
(I) in an initial noncondensed state ($\theta^0=0$, $t$=0), and  
(II) in a condensed state in chemical equilibrium 
($\theta=\theta^{\rm eq}$,\footnote{
The superscript `{\rm eq}' denotes the equilibrium value. }
$t\rightarrow \infty$). 

\noindent (i) {\it Initial noncondensed state (I) } 

\begin{figure}[t]
\centerline{
\epsfxsize=0.5\textwidth
\epsffile{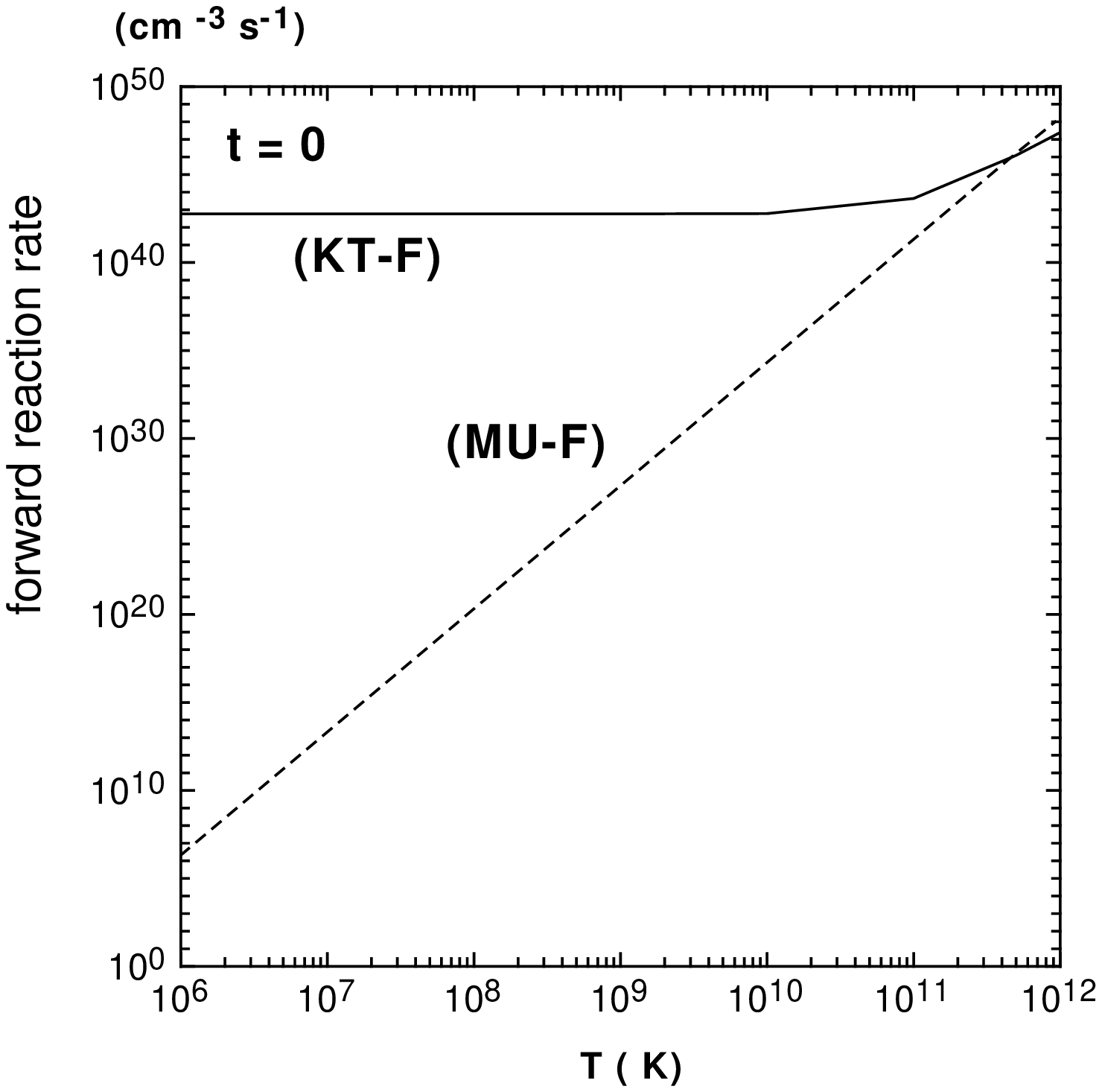}}
\caption{\footnotesize  Temperature dependence  
of the forward KT reaction rate $\Gamma^{\rm (KT-F)}(\xi^{\rm
(KT)}(0),T)$  at an initial  noncondensed stage [case (I)] for $n_B$=0.55
fm$^{-3}$ (solid line).  For comparison, the plot for the forward MU
reaction rate is shown by the dashed line. }
\label{fig2}
\end{figure}

\begin{figure}[tt]\noindent
\begin{minipage}[l]{0.5\textwidth}
\centerline{
\epsfxsize=0.90\textwidth
\epsffile{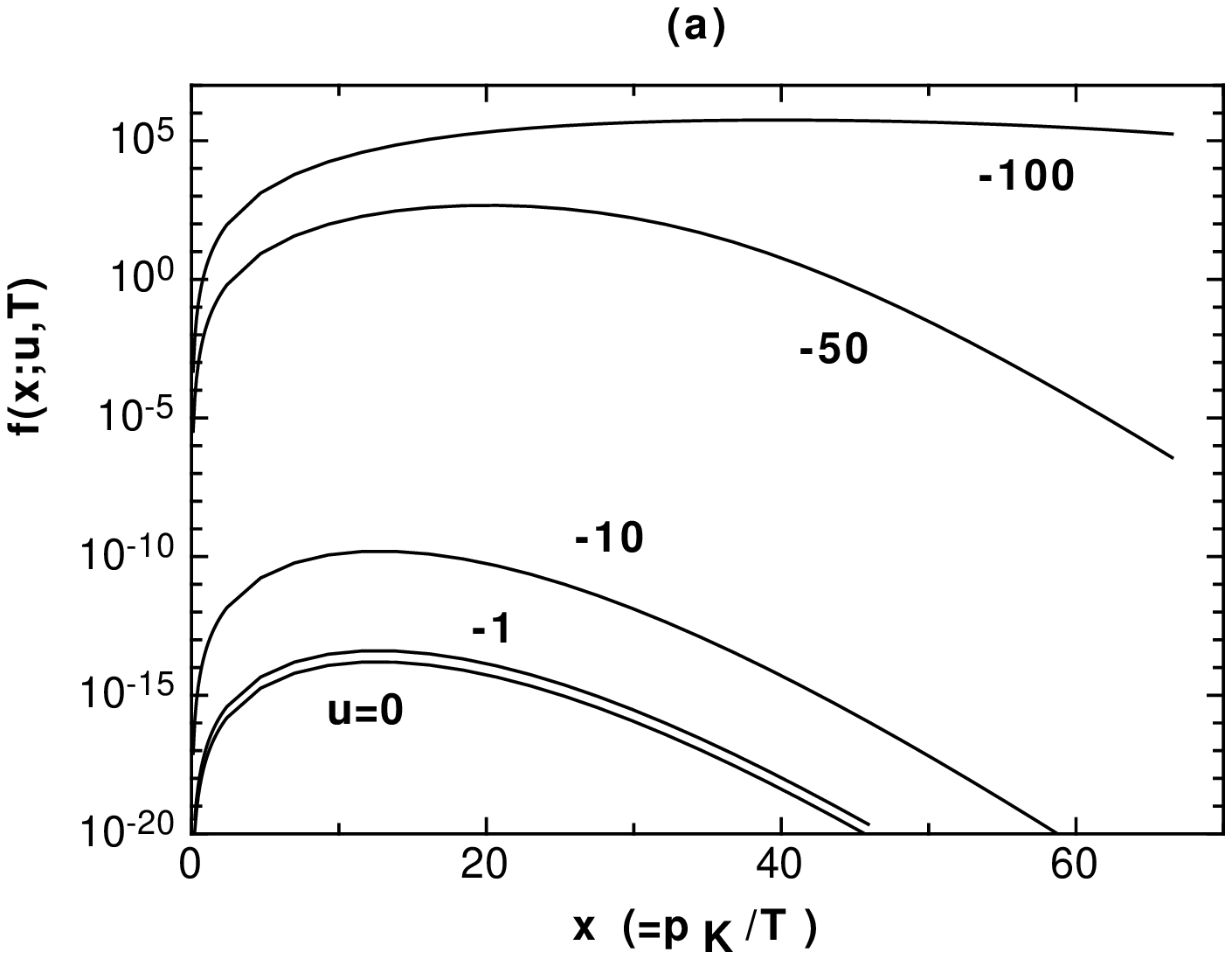}}
\end{minipage}~
\begin{minipage}[r]{0.5\textwidth}
\centerline{
\epsfxsize=\textwidth
\epsffile{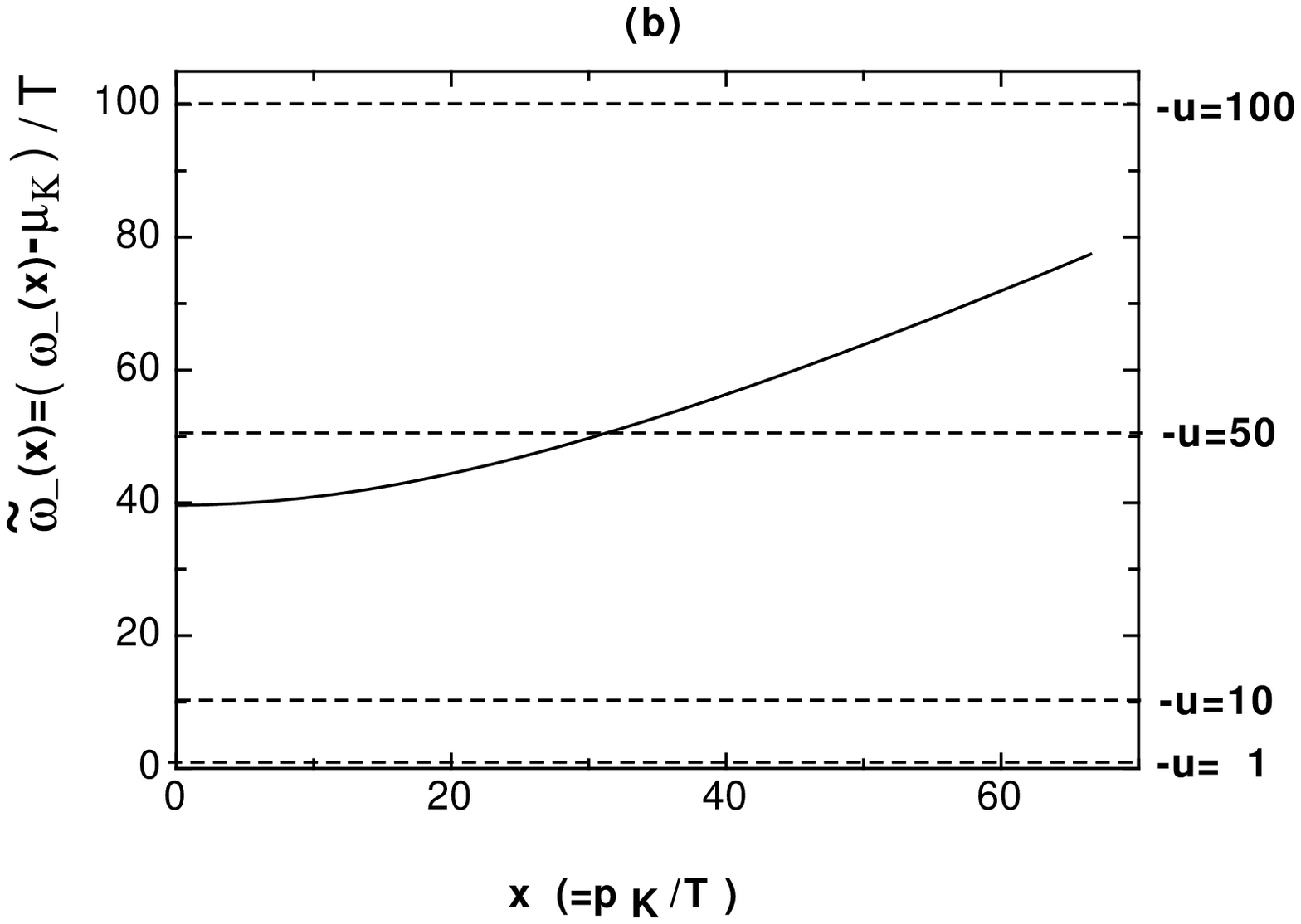}}
\end{minipage}
\caption{\footnotesize  (a) Function $f(x;u,T)$ 
in the integrand $I^{(\rm KT) }(u,T)$  for the KT-F reaction rate 
as a function of 
$x (=|{\bf p_K}|/T)$ for several values of $u$  
at an initial noncondensed stage (case (I)). 
The result is for $n_B$=0.55 fm$^{-3}$ and $T=1.0\times 10^{11}$ K. \\  
(b) $\widetilde{\omega_-}(x) [=(\omega_-(x)-\mu_K)/T ]$, the kaon
excitation energy normalized by the temperature 
under the same condition as in Fig.3(a).}
\label{fig3} 
\end{figure}

First we consider a case with the baryon number density $n_B$
=0.55 fm$^{-3}$,  which  is just above the critical density $n_B^C$. In this 
case, we take $\mu_K^0=-b^0=-139$ MeV, 
and the initial proton-mixing ratio, $x_p^0\equiv n_p^0/n_B$=0.14. 
The initial values of the physical quantities are listed in Table I 
for the two baryon number densities. 
In Fig.\ref{fig2}, the temperature dependence of 
$\Gamma^{\rm (KT-F)}(\xi^{\rm (KT)}(0),T)$ 
is shown by the solid line. 
For comparison, the reaction rate for the forward MU process (MU-F)   
is shown by the dashed line 
(see Appendix A for the expression for the MU reaction rate). 
The MU-F reaction rate is taken for $\beta$ equilibrium, i.e., $\xi^{({\rm
MU})}(0)$=0.  Note that there is no condensate  yet 
at $t=0$, so that the KU reactions cannot proceed. 
One can see that the magnitude of $\Gamma^{\rm (KT-F)}
(\xi^{\rm (KT)}(0),T)$ is large  and dominates over that for MU,
$\Gamma^{\rm (MU-F)}(\xi^{\rm (MU)}(0)=0,T)$,  over most cases
except at very high temperature,
$T\gtrsim 5\times 10^{11}$  K. The large reaction rate for KT is 
due to the large deviation of the system from chemical equilibrium. For
example,  the value of $\xi^{\rm (KT)}(0)$ is $-4.6\times 10^3/T_9 \ll 
0$ at a relevant temperature $T\lesssim 10^{11}$ K. 
To see the $\xi^{\rm (KT)}$-dependence of $\Gamma^{\rm (KT-F)}$, we 
show the integrand $f(x;u,T)$ of $I^{\rm (KT-F)}(u,T)$ [ Eq.(\ref{ikt}) ]
in Fig.\ref{fig3} (a), as a function of 
$x$(=$|{\bf p_K}|/T$) with the temperature fixed at $T=1\times 10^{11}$ K. 
Several values of $u$ are taken for comparison. 
In Fig.\ref{fig3} (b), the normalized kaon excitation energy 
$\widetilde{\omega_-}$ [=$(\omega_-(x)-\mu_K)/T$ ] is also shown.  
The value of $f(x;u,T)$ for $u=0$, in which case the normal neutron-star
matter would be in chemical equilibrium without a kaon condensate, is much
reduced, as is expected,  because the energy gap measured from the 
chemical potential [$\widetilde{\omega_-}(x=0)=
(\omega_-(0)-\mu_K)/T >
0$] reduces the value of the Bose-Einstein distribution function  [cf.
Fig.\ref{fig3} (b) ]. However, as the value of 
$u$ changes from zero to $-100$, the maximum value of $f(x;u,T)$, 
$f_{\rm max}(u,T)$, becomes larger 
by more than ten orders of magnitude. 
One can also see from Fig.\ref{fig3} (a) that 
the kaon momentum ($x$) at $f_{\rm max}(u,T)$ becomes larger 
as $|u|$ increases, and that a wider range of kaon momenta 
contribute to the integral $I^{\rm (KT-F)}(u,T)$.  
For the highly nonequilibrium case $|u|\gtrsim 50$, 
the main contribution to the reaction rate is 
associated with the hard thermal kaons with large momenta, 
rather  than the soft kaons with
$\widetilde{\omega_-}(x) -\widetilde{\omega_-}(0)= O(1)$, 
as seen in Fig.\ref{fig3} (b).  

At lower temperatures, $T\lesssim 10^{10}$
K, the temperature dependence of the reaction rate 
$\Gamma^{\rm (KT-F)}(\xi^{\rm (KT)}(0),T)$ is weak, as seen in
Fig.\ref{fig2}.   In order to  look into the details of this behavior,  
we show, in Fig.4,   the integrand  
$f(x; u=\xi^{\rm (KT)}(0),T)$  of $I^{\rm ( KT-F)}$ as a function of $x$ 
with the input parameters $\mu_K^0$ and $x_p^0$  for $n_B$=0.55
fm$^{-3}$  and for several temperatures, $T=10^8-10^{13}$ K. 
One can see that, for $T\lesssim 10^{10}$ K,  
$f(x; u=\xi^{\rm (KT)}(0),T)$
is nearly proportional to $x^4$ for $x\lesssim x_M
\equiv |{\bf p}_{K,M}|/T$, where $|{\bf p}_{K,M}|$ is the kaon momentum
corresponding to a maximum of $f$. Above $x=x_M$, 
 the integrand $f$ decreases rapidly with $x$. 
This $x$-dependence of the integrand 
$f(x; u=\xi^{\rm (KT)}(0),T)$ can be derived as follows:
With the input parameters $\mu_K^0$ and
$x_p^0$,  one finds $\widetilde{\omega_-}(x)\leq\widetilde{
\omega_-}(x_M)=4.2\times 10^3/T_9$ for $x\leq x_M$, and 
$\xi^{\rm(KT)}(0)=-4.6\times 10^3/T_9$, so that 
$\widetilde{\omega_-}(x)+\xi^{\rm(KT)}(0)\lesssim 
-4\times 10^2/T_9$ for $x\leq x_M$. Thus, one can  show from
Eq.(\ref{integrandf}) that  
$\displaystyle f(x; u=\xi^{\rm (KT)}(0),T)
\sim \frac{1}{6}\frac{x^4}{(\widetilde{\omega_-}(x)+\mu_K^0/T)^3}
|\widetilde{\omega_-}(x)+\xi^{\rm (KT)}(0)|^3 \propto x^4$, 
noting that 
$ |\widetilde{\omega_-}(x)+\xi^{\rm (KT)}(0)|^2\gg 4\pi^2 $ for
$T\lesssim 10^{10}$ K, and that the $x$-dependence of 
$\widetilde{\omega_-}(x)$ is negligible for  $x\lesssim x_M$. 
At a certain value of $x$ just beyond $x_M$, 
$\widetilde{\omega_-}(x)+\xi^{\rm(KT)}(0)$ becomes positive, and 
 the integrand
$f(x; u=\xi^{\rm (KT)}(0),T)$ decreases rapidly with $x$ due to the
exponential factor in (\ref{integrandf}), which originates from 
the statistical factor in the phase-space integrals of the reaction rate. 
Hence one obtains $\displaystyle I^{\rm ( KT-F)}\sim
\int_0^{x_M}fdx\propto x_M^5$. It is to be noted that 
the kaons which have the most dominant contribution to the KT 
reactions have momenta  around a hundred  MeV. 
For example, the value of the 
momentum $|{\bf p}_{K,M}|$ is $ |{\bf p}_{K,M}| 
\simeq $120 MeV for $n_B=0.55$ fm$^{-3}$. 
This kinematical  range for the kaons is almost  
{\it independent} of the temperature as far as the temperature is low 
such that $T\lesssim 10^{10}$ K(=0.86 MeV).  
Therefore, $x_M(=|{\bf p}_{K,M}|/T)$ is simply proportional to
$T^{-1}$,  and  $I^{\rm ( KT-F)}\propto T^{-5}$. Thus one obtains 
$\Gamma^{\rm (KT-F)}(\xi^{\rm (KT)}(0),T)=O(T^0)$  from
Eq.(\ref{frrfktb}).  

In the high temperature case, $T\gtrsim 10^{12}$ K, on the other hand, 
the large energy of thermally excited kaons [$T \gtrsim O $(100
MeV) ] is  comparable to the  energy gap, $\omega_-({\bf
p}_K=0)-\mu_K$, the latter of which is estimated to be  340 MeV for
$n_B$=0.55 fm$^{-3}$. Thereby the soft kaons  with energy less
than the energy corresponding to 
$\widetilde{\omega_-}(x)=O(1)$ contribute to the relevant reactions. 
For $T\gtrsim 10^{12}$ K, a wide range of  momenta of the thermal
kaons give the main contribution to the  reaction rate, 
and the value of $|{\bf p}_{K,M}|$  depends sensitively 
on the temperature, as one can see in Fig.4. 

\noindent (ii) {\it Condensed state in chemical equilibrium (II) }
 
For $n_B$=0.55fm$^{-3}$, the input parameters in the  
equilibrated kaon-condensed phase are taken to be  
$\theta^{\rm eq}$ =0.48,  
$\mu_K^{\rm eq}$=203 MeV, and $x_p^{\rm eq}$=0.23. 
The temperature dependence of the forward reaction rate 
$\Gamma^{\rm (KT-F)}(\xi^{\rm (KT)},T)$ with $\xi^{\rm (KT)}= 0$ 
is shown in Fig.\ref{fig5}\ by the solid line 
for the baryon number density $n_B$=0.55fm$^{-3}$, 
together with a plot of the forward 
KU process (KU-F) ( the MU-F process )  
shown by the dotted line (dashed line).  
(see Appendix A for the expression for $\Gamma^{\rm (KU-F)}$.)

In order to discuss the kinematics for kaons in the KT 
reaction quantitatively, we show, in Fig.\ref{fig6}, $f(x; u=\xi^{\rm
(KT)}=0,T)$ and 
$\widetilde{\omega_-}(x)$ for $n_B$=0.55 fm$^{-3}$ and $T=10^9$,
$10^{10}$, and $10^{11}$ K. 
One can see that the maximum of 
$f(x; u=\xi^{\rm (KT)}=0,T)$ corresponds to 
$\widetilde{\omega_-}(x)=a=O(1)$, where $a=2-3$, 
i.e.,  {\it soft thermal kaons} with thermal energies produce
 a dominant contribution to the KT reaction in the condensed phase.
\footnote{The importance of thermal loops 
in the soft (Goldstone) mode for  the phase diagram 
and EOS has been also emphasized in ref.\cite{ty98}.}
 This feature holds for all temperatures. 
By expanding 
$\widetilde{\omega_-}(x)$ with respect to the  kaon momentum 
$|{\bf p}_K|$ by the use of the classical
field equation, Eq.(\ref{fieldeq}), 
one obtains the momentum 
$|{\bf p}_{K,M}|$ corresponding to the maximum of the integrand, $f_{\rm
max}$,  as
\begin{equation}
|{\bf p}_{K,M}|=\sqrt{2a(b^{\rm eq} +\mu_K^{\rm eq}
\cos\theta^{\rm eq})T}\propto T^{1/2} 
\label{appropk}
\end{equation}
for the temperatures $T\lesssim 10^{11}$ K, or $x_M=
|{\bf p}_{K,M}|/T\propto T^{-1/2}$. The temperature-dependence of  
$f_{\rm max}$ is then written roughly as 

$\displaystyle f_{\rm max}\sim 
\frac{1}{6}\frac{[2a(b^{\rm eq} +\mu_K^{\rm eq}\cos\theta^{\rm
eq})]^2}{T^2(a+\mu_K^{\rm eq}/T)^3}\frac{a}{1-e^{-a}}
\frac{(a^2+4\pi^2)}{e^a-1}$.    In the
low temperature limit, $\mu_K^{\rm eq}/T\gg a$,
which  gives $f_{\rm max}\propto T$. 
In this case,  $\displaystyle I^{\rm KT-F}(\xi^{\rm (KT)}=0,T)\sim 
f_{\rm max}\Delta x\propto T^{1/2}$, where $\Delta x(=\Delta p_K/T)$ 
is the dominant integration  range around $x_M$. 
Hence one finds $\Gamma^{\rm (KT-F)}(\xi^{\rm (KT)}=0,T)\propto
T^{5.5}$.  

At temperatures below $\sim10^8$ K, the reaction rate for KT, 
$\Gamma^{\rm (KT-F)}(\xi^{\rm (KT)}=0,T)$, is less 
than that for KU, $\Gamma^{\rm (KU-F)}(\xi^{\rm (KU)}=0,T)$. 
Nevertheless, since the temperature dependence of the KT process 
($\propto T^{5.5}$) is more pronounced than that for the KU process
($\propto T^5$), the reaction rate $\Gamma^{\rm (KT-F)}(\xi^{\rm 
(KT)}=0,T)$ 
increases more rapidly as a function of the temperature than
$\Gamma^{\rm (KU-F)}(\xi^{\rm (KU)}=0,T)$,  and the former exceeds
the latter for
$T\gtrsim 10^8$ K.  At temperatures $T\gtrsim 10^9$ K,
$\Gamma^{\rm (KT-F)}(\xi^{\rm  (KT)}=0,T)$ is larger than
$\Gamma^{\rm (KU-F)}(\xi^{\rm (KU)}=0,T)$ by  one or two orders of
magnitude, and at temperatures below $\sim 4\times 10^{11}$ K, 
$\Gamma^{\rm (KT-F)}(\xi^{\rm (KT)}=0,T)$  is also larger than 
$\Gamma^{\rm (MU-F)}(\xi^{\rm (MU)}=0,T)$ . 

\noindent (iii) {\it Comparison between cases (I) and (II) }

Next we compare the reaction rate for KT in the case (I) with that in 
the case (II) (cf. Figs.\ref{fig2} and \ref{fig5}\ ). 
The former is larger than the latter for all the temperatures. 
For example, the former is large in magnitude by a factor
$\sim 10^6$ at $T=10^{10}$ K, and this factor becomes  much more 
significant at lower  temperatures,  
where hard thermal kaons with $\widetilde{
\omega_-}(x)\gg 1$ take part in an enhancement 
of the KT reactions in the nonequilibrium state.  
As the temperature increases,
the ratio, 
$\Gamma^{\rm (KT-F)}[{\rm case \ (II)}]/\Gamma^{\rm
(KT-F)}[{\rm case \ (I)}]$, becomes smaller, and tends to have a 
 $T$ dependence similar to each other at very high temperatures, 
$T\gtrsim 10^{12}$ K,
where soft kaons with energies $\widetilde{\omega_-}(x)=O(1)$  
become responsible for the reactions KT for both (I) and (II). 

\noindent (iv) {\it Comparison at different densities}

Finally, in Fig.7, we show the temperature dependence of the 
relevant forward reaction rates for $n_B$=0.70 fm$^{-3}$. At this density, 
we have the fully-developed kaon-condensed phase 
after chemical equilibrium is attained. The values for the parameters are 
estimated to be $\mu_K^0=-b^0=-180$ MeV and $x_p^0$=0.16 for the
case  (I), and $\theta^{\rm eq}$ =0.91,  $\mu_K^{\rm eq}$=114 MeV, and 
$x_p^{\rm eq}$=0.39 for the case (II).  
 Figure \ref{fig7} (a) [ Fig.\ref{fig7} (b) ] is for the  case (I) 
[case (II) ].
The qualitative features are the same as those at a lower density 
$n_B$=0.55 fm$^{-3}$, although there is a significant increase in the
magnitude  of the KT reaction rate at the same temperature as
compared with the lower density case of $n_B$=0.55 fm$^{-3}$ [ cf.
Fig.\ref{fig2} for (I) and Fig.\ref{fig5} for (II) ].  

In the initial noncondensed case (I), the value of
$\xi^{\rm (KT)}(0)$  is further reduced to 
$\xi^{\rm (KT)}(0) =-5.5\times10^3/T_9$ with increase in density,
because  the decrease in
$\mu_K^0$ and the increase in $\mu_e^0$ enlarge the value of 
$|\xi^{\rm (KT)}(0)|$(=$|\mu_e^0-\mu_K^0|/T$). 
 This  difference in $\xi^{\rm (KT)}(0)$ becomes important in
determining the magnitude of the KT reaction rate for temperatures 
$T\lesssim 10^{11}$ K. For example, the ratio  
of the reaction rates at different densities, 
$r({\rm I})=\Gamma^{\rm (KT-F)}(0.70{\rm fm}^{-3})/\Gamma^{\rm
(KT-F)}(0.55{\rm fm}^{-3})$, becomes $r({\rm I}) \sim 10^3$ 
for $T\lesssim 10^{10}$ K.  For temperatures as high as 
$T\gtrsim 5\times 10^{11}$ K, the difference 
in $\xi^{\rm (KT)}(0) $ becomes less significant because 
$|\xi^{\rm (KT)}(0)|\lesssim 10 $  at both densities, so that the ratio 
 becomes small: 
$r({\rm I})=\Gamma^{\rm (KT-F)}(0.70 {\rm fm}^{-3})/\Gamma^{\rm
(KT-F)}(0.55 {\rm fm}^{-3}) 
\lesssim  10$ for $T\gtrsim 5\times 10^{11}$ K.

On the other hand, in the case of the condensed state in chemical 
equilibrium (II), the ratio $r({\rm II})=
\Gamma^{\rm (KT-F)}(0.70{\rm
fm}^{-3})/\Gamma^{\rm (KT-F)}(0.55{\rm fm}^{-3}) \sim 3$ at low
temperatures 
$T\lesssim 10^{10}$ K, while $r({\rm II})$ becomes less than 2 at high
temperatures 
$T\gtrsim 5\times 10^{11}$ K. 
This enhancement with increase in density 
mainly stems from the reduction 
in the kaon chemical potential $\mu_K^{\rm eq}$, which appears 
in the denominator of the function $f(x; u=\xi^{(\rm KT)}=0,T)$: 
In the kaon-condensed phase, $\mu_K^{\rm eq}$ decreases 
monotonically as the density increases\cite{fmtt94} such that 
$\mu_K^{\rm eq}$=203 MeV$\rightarrow$ 114 MeV 
as $n_B$=0.55 fm$^{-3}$$\rightarrow$ 0.70 fm$^{-3}$. 
At low temperatures, the term 
$\mu_K^{\rm eq}/T$ is larger than 
$\widetilde {\omega_-}(x)$(=$O(1)$)  in the factor 
$1/(\widetilde{\omega_-}(x)+\mu_K^{\rm eq}/T)^3$,  so that 
$\displaystyle r({\rm II})$ is mainly determined by 
$\big\lbrack\mu_K^{\rm eq}(0.55 {\rm fm}^{-3})
/ \mu_K^{\rm eq}(0.70 {\rm fm}^{-3})\big\rbrack^3$.  
At high temperatures, the term
$\mu_K^{\rm eq}/T$ is less significant,  so that the difference of the
integral $I^{\rm (KT-F)}(\xi^{({\rm KT})}=0,T)$ between  the different
densities is less marked. 

\section{Summary and concluding remarks}
\label{sec:summary}

We have calculated the reaction rates for 
the thermal kaon (KT) process.  We based on chiral symmetry 
as a guiding principle in obtaining the excitation energy of the kaons 
and the transition matrix element for the reaction rates.  
It has been shown that the reaction rate is larger than 
those of the kaon-induced Urca (KU) 
and the modified Urca (MU) reactions for the relevant temperatures 
and baryon number densities which may be realized 
in the early hot stage of neutron stars. 
The KT process is dominant not only in the case of  
 the noncondensed state, which is in a highly nonequilibrium state 
($|\xi^{\rm KT}(t)|\gg 1$), but also in the case of 
 the kaon-condensed state  in chemical equilibrium 
($|\xi^{\rm KT}(t)|=0$).  In the noncondensed state, 
where there is a gap
between the minimum  excitation energy 
and the kaon chemical potential, it is mainly 
hard thermal  kaons with large momenta,
 rather than soft thermal kaons,  
which contribute to the reaction rate. On the other hand, 
in the condensed state, the soft kaon mode 
[$\omega_-({\bf p}_K) -\mu_K=O(T)$] , which reflects 
the spontaneously broken $V$-spin symmetry,  
contributes to the reaction rate. 
We have seen that the hard and soft kaons contribute differently, 
which  results in a different temperature dependence for the reaction
rates in the noncondensed and condensed states. 

The KT reactions are found to be dominant 
throughout the nonequilibrium process, and may control the 
characteristic time scales, such as those for the onset of 
condensation and its subsequent buildup.  Dynamical evolution 
of the kaon-condensed state can be treated  by pursuing temporal 
changes in the condensation by way of a set of kinetic equations, 
where the chemical species change 
through the nonequilibrium weak reactions. 
This subject will be discussed in detail elsewhere\cite{mti98-2}.  

In addition, there are several astrophysical implications 
of the present work. 
First, Brown and Bethe\cite{bb94} have proposed a scenario 
in which low-mass black holes are formed in stellar collapse 
using a very soft equation of state  due to kaon condensation. 
Following this scenario,  Baumgarte et al. made a dynamical simulation of the 
delayed collapse of a hot neutron star to a black hole\cite{bst96}. 
They utilized the nuclear EOS with kaon condensation 
in the equilibrium configuration at $T=0$. 
However, there might be a time lag for the appearance and 
growth of a condensate due to the nonequilibrium weak reactions, 
even when the density exceeds the critical density for the condensation.  
Hence one needs to treat the nonequilibrium processes  
for a careful consideration of the dynamical evolution of an initially  
normal neutron star to a kaon-condensed star.  Second, 
nonequilibrium weak reactions may also affect the 
stability of neutron stars near the maximum mass  
with respect to a perturbation from 
their static configurations\cite{ghg95}.  
For normal nuclear matter, the $\beta$ processes (\ref{mu}) 
have been taken into account 
as the most important weak reactions\cite{ghg95}.    
For kaon-condensed matter, the KT process is expected 
to be the most relevant reactions for stability. Third, 
as another dynamical property related to neutron stars, 
 mechanisms for the dissipation of the 
vibrational energy of neutron stars induced  
by nonequilibrium weak reactions have been 
discussed for several phases of hadronic matter\cite{cls90}.  
The associated bulk viscosity determines the damping time scale  
for radial oscillations of neutron stars. 
In a kaon condensate, the KT reactions may be most 
effective for dissipation, and may have a major contribution 
to the damping of the radial oscillations of kaon-condensed 
stars.   

\section{Acknowledgements}
The authors wish to thank Professor R.T.Deck for comments on the
manuscript. 
 One of the authors (T.M.) is indebted for the Grant-in-Aid of 
Chiba Institute of Technology (C.I.T). Part of the numerical calculations 
was performed by the use of the DEC Alpha Server 4100 System in 
C.I.T. This material is based upon work supported in part by the 
National Science Foundation through the Theoretical Physics Program 
under Grant Nos.PHY9008475 and PHY9722138, and by the Japanese 
Grant-in-Aid for Scientific Research Fund of the Ministry of Education, 
Science, Sports and Culture (08640369, 11640272). 

\appendix
\section{Other weak reactions}

For comparison, we here list other relevant 
weak reactions KU and MU, which may be operative in  
the nonequilibrium process in kaon condensation. 

The KU reactions are mediated by the last term (the ``commutator''
contribution) in Eq.(\ref{tjh}). 
The reaction rate for the forward KU process (\ref{fku}), is given 
as  
\begin{eqnarray}
\Gamma^{({\rm KU-F})}(\xi^{({\rm KU})},T) 
&=&\frac{G_F^2}{64\pi^5}\sin^2\theta_C\sin^2\theta\Big\{
10+3(g_A^2+9{\widetilde g_A}^2)\Big\}m_N^{\ast 
2}\mu_eT^5I_2(\xi^{({\rm KU})}) \cr
&=&(6.6\times 10^{29})\bigg({m^\ast_N \over m_N}\bigg)^2
{\mu_e\over m_\pi}\sin^2\theta T_9^5I_2(\xi^{({\rm KU})}) \
({\rm cm}^{-3}\cdot {\rm s}^{-1}) 
\label{rrfku}
\end{eqnarray}
where $\widetilde g_A=F-\frac{1}{3}D$=0.15 with $F+D=g_A=1.25$ and 
$D/(D+F)=0.658$\cite{fmtt94}, 
$\displaystyle I_2(u)\equiv \int_0 ^\infty dx x^2[\pi^2+(x+u)^2]/
( 1+\exp(x+u))$ and   
$\xi^{\rm (KU)} \equiv (\mu_e -\mu_K )/T$.   

For the MU reaction, we refer to Haensel's result\cite{h92} which 
is based on \cite{fm79}.  
Noting that the matrix elements are slightly modified in
the presence of the kaon-condensate by an
additional factor of $\cos^2(\theta/2)$ coming from the first term in 
the first curly brackets for the 
isospin-changing strangeness-conserving current in 
(\ref{tjh}), one obtains
\begin{equation}
\Gamma^{({\rm MU-F})}(\xi^{({\rm MU})},T)
=(5.9\times 10^{23})\bigg({n_e\over n_0}\bigg)^{1/3}
\cos^2{\theta\over 2}T_9^7 J_2(\xi^{({\rm MU})})\ 
({\rm cm}^{-3}\cdot {\rm s}^{-1}) \ ,
\label{rrfmu}
\end{equation}
where 
$\displaystyle J_2(u)\equiv \int_0 ^\infty dx
x^2[9\pi^4+10\pi^2(x+u)^2+(x+u)^4]
/( 1+\exp(x+u))$, and 
$ \xi^{\rm (MU)}\equiv (\mu_p+\mu_e-\mu_n)/T$. 

For the backward processes, one can see  
$\Gamma^{\rm (KU-B)}
=\Gamma^{({\rm KU-F})}(-\xi^{({\rm KU})}) $, 
$\Gamma^{\rm (MU-B)}
=\Gamma^{\rm (MU-F)}(-\xi^{({\rm MU})}) $  
within the low-temperature approximation. 
It is to be noted that the relation 
between the forward and backward reaction rates for KT [
(\ref{brrkt}) ] is different from that for KU or MU 
due to the appearance of the Bose-Einstein distribution function 
in the phase-space integrals for KT.

\begin{table}
 \caption{\footnotesize Input quantities for the initial noncondensed
state ($t=0$) and for the  chemical equilibrated  state ($t\rightarrow
\infty$). All the values are estimated at $T$=0. The former (the latter)
quantities are denoted by the superscript `0' ( `eq' ). }
\label{tabcond} 
\begin{tabular}{c || c c | c c c}
$n_{\rm B}$({\rm fm}$^{-3}$) &$\mu_K^0 $ ( MeV ) & $x_p^0$  & 
$\theta^{\rm eq}$ (${\rm rad}$) & $\mu_K^{\rm eq} $ ( MeV )
& $x_p^{\rm eq}$ \\
\hline\hline 
 0.55 & $-139$ & 0.14 & 0.48 & 203  & 0.23 \\
\hline 
0.77 &  $-180$ & 0.16 & 0.91 & 114 & 0.39 \\ 
\end{tabular}
\end{table}

\vspace{1.0cm}

\vspace{1.0cm}

\vspace{1.0cm}

\begin{figure}
\centerline{
\epsfxsize=0.5\textwidth\epsffile{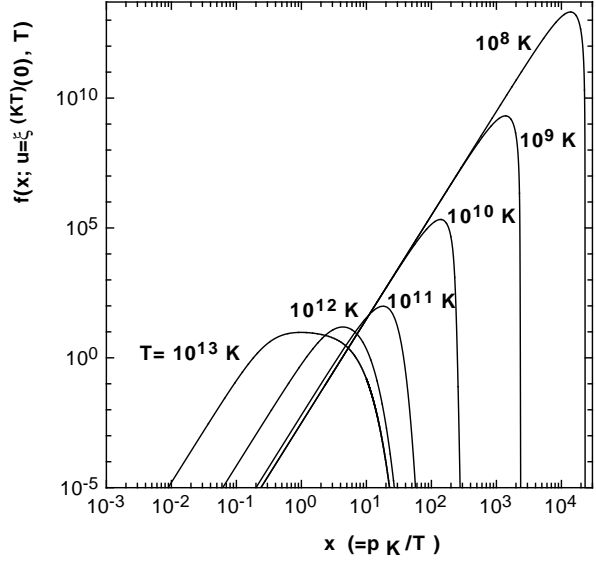}}
\label{fig4}
\caption{\footnotesize  The integrand $f(x; u=\xi^{\rm KT}(0),T)$ 
  of $I^{\rm (KT-F)}$ as a function of $x$  with the input parameters
$\mu_K^0$ and
$x_p^0$,  for $n_B$=0.55 fm$^{-3}$ and for several temperatures, 
$T=10^8-10^{13}$ K.}
\end{figure}
\vspace{1.0cm}

\begin{figure}[t]
\centerline{
\epsfxsize=0.5\textwidth\epsffile{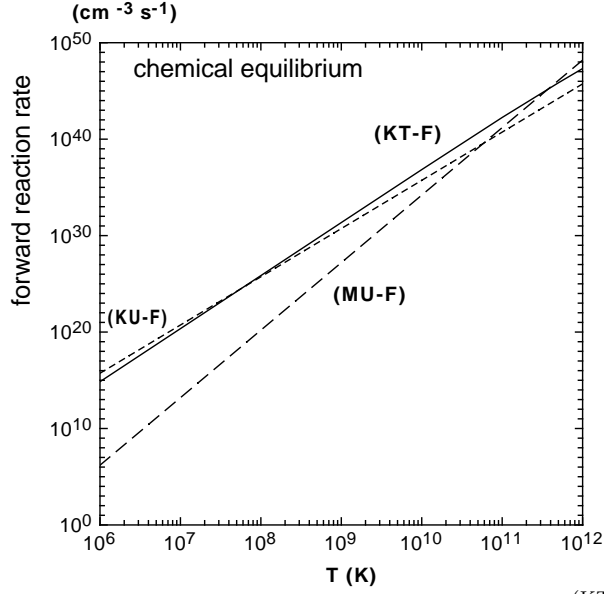}}
\caption{\footnotesize Temperature dependence of the forward KT 
reaction rate 
$\Gamma^{\rm (KT-F)}(\xi^{\rm (KT)}=0,T)$ in the kaon-condensed 
phase in chemical equilibrium (II) for $n_B$=0.55fm$^{-3}$ (solid line).
For comparison, the reaction rate for the KU process
(the MU process) is shown with a dotted line (dashed line).}
\label{fig5}
\end{figure}
\vspace{1.0cm}

\begin{figure}[tt]
\begin{minipage}[l]{0.5\textwidth} 
\centerline{
\epsfxsize=\textwidth
\epsffile{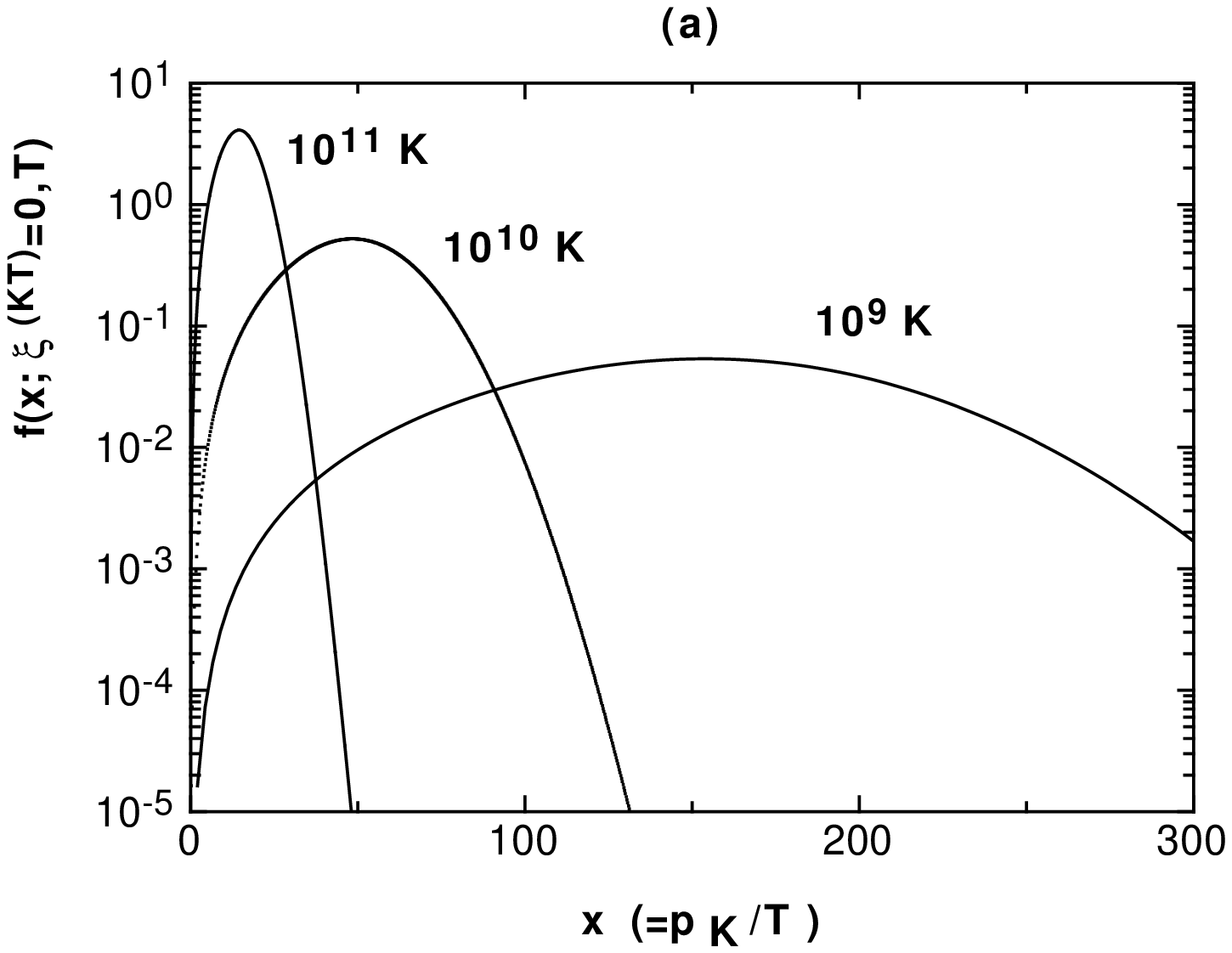}}
\end{minipage}~
\begin{minipage}[r]{0.5\textwidth}
\centerline{
\epsfxsize=\textwidth
\epsffile{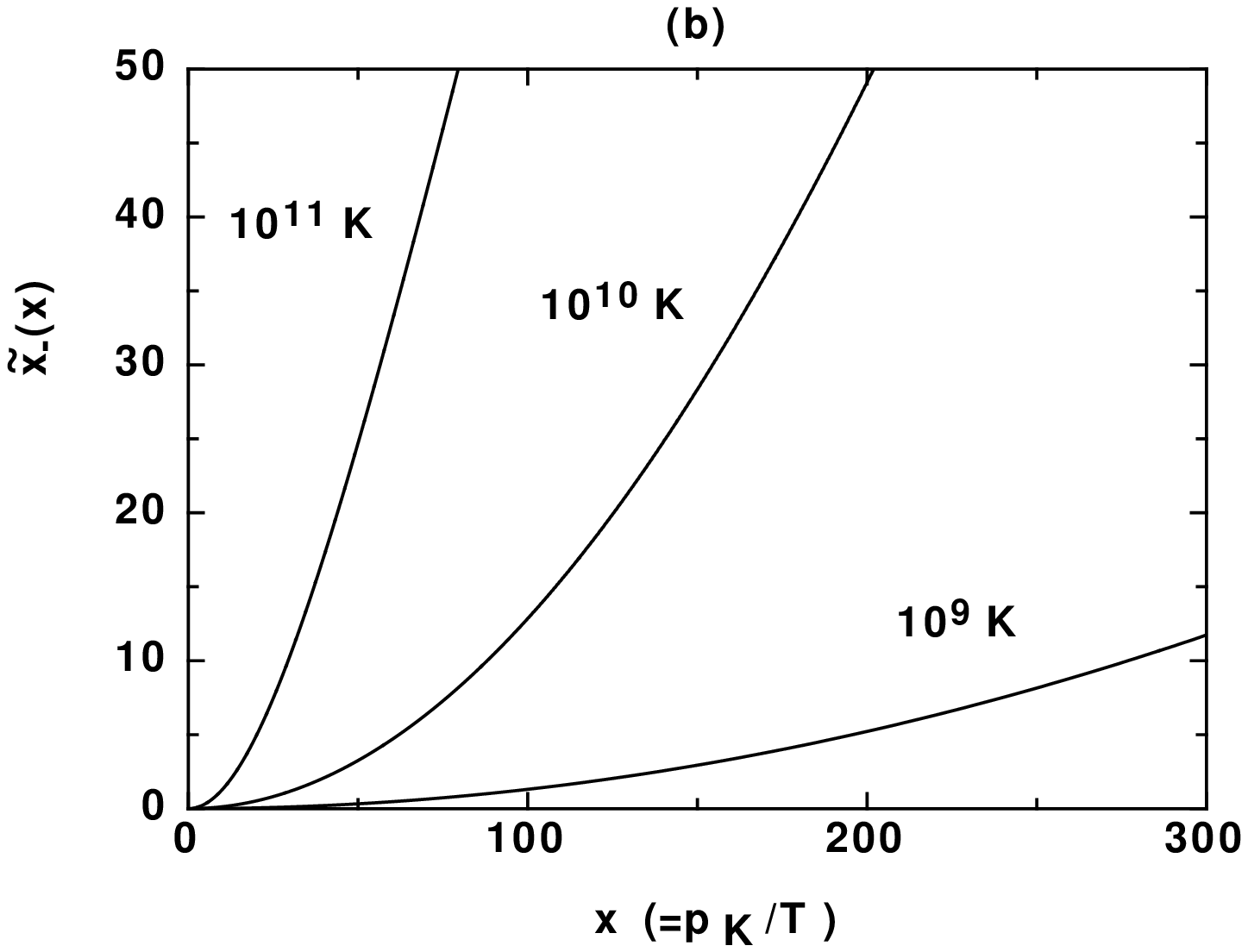}}
\end{minipage}
\caption{\footnotesize  (a) Function $f(x; u=\xi^{(\rm KT)}=0,T)$ 
at the equilibrated $K^-$-condensed stage (II) 
as a function of $x$ for $n_B$=0.55fm$^{-3}$ and several 
temperatures. \\ 
(b) $\widetilde{\omega_-}(x)$ as a function of $x$ 
under the same condition as Fig.6(a).}  
\label{fig6}
\end{figure} 
\vspace{1.0cm}

\begin{figure}[tt]
\begin{minipage}[l]{0.4\textwidth} 
\centerline{
\epsfxsize=\textwidth
\epsffile{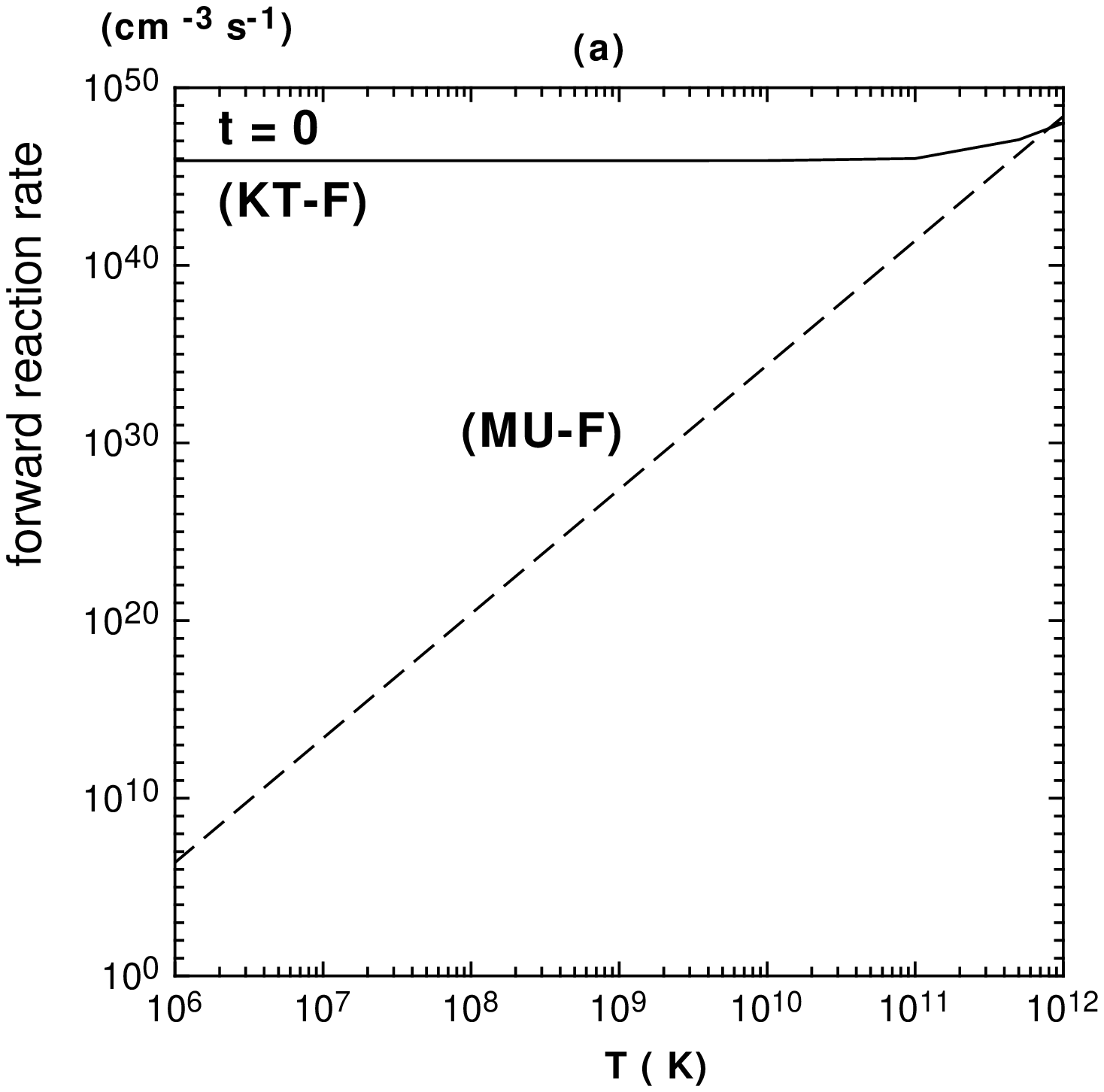}}
\end{minipage}~\hspace{1.0cm}
\begin{minipage}[r]{0.4\textwidth}
\centerline{
\epsfxsize=\textwidth
\epsffile{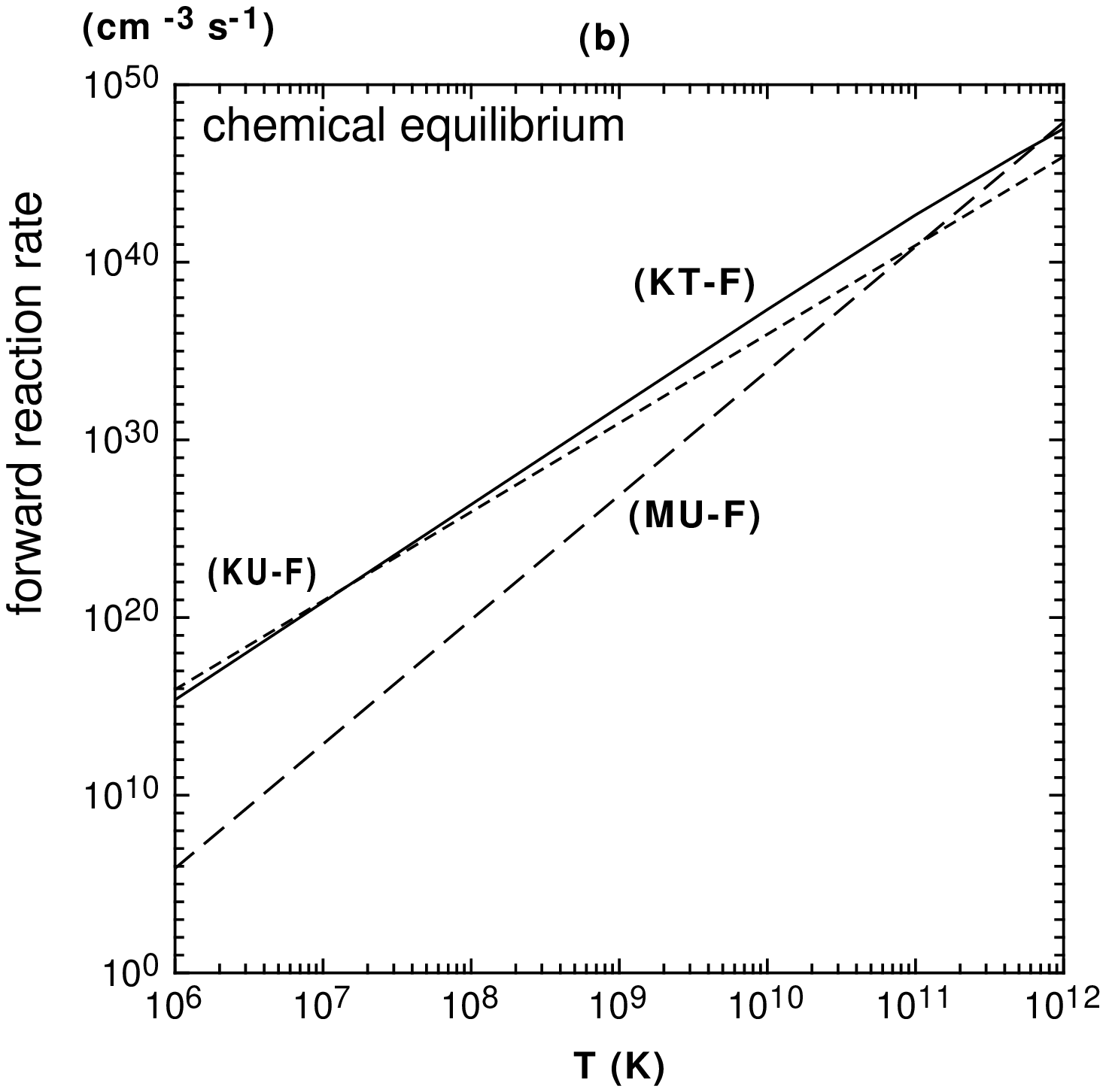}}
\end{minipage}
\caption{\footnotesize 
(a) Temperature dependence of the forward KT reaction 
rate $\Gamma^{\rm (KT-F)}(\xi^{\rm (KT)}(0),T)$  at an initial 
noncondensed stage [case (I) ] for $n_B$=0.70fm$^{-3}$ (solid line). 
For comparison, the plot for the forward MU reaction is shown 
by the dashed line. \\ 
(b) Temperature dependence of the forward KT reaction rate 
$\Gamma^{\rm (KT-F)}(\xi^{\rm (KT)}=0,T)$ in the kaon-condensed 
phase in chemical equilibrium [case (II) ] for $n_B$=0.70 fm$^{-3}$ (solid
line). For comparison, the rate for the KU process 
(the MU process) is shown by the dotted line (dashed line).}
\label{fig7}
\end{figure}

\end{document}